\pdfoutput=1
\documentclass[11pt,twoside,a4paper,cmspaper,final,collab]{cms-tdr}
\def\svnVersion{b7254f4}\def\svnDate{2026/07/23}\def\cmsCernNoTag{CERN-EP-2025-251}\def\cmsCernDate{\today}\def\cmsMessage{Published in the Journal of High Energy Physics as \href{http://dx.doi.org/10.1007/JHEP07(2026)119}{\doi{10.1007/JHEP07(2026)119}.}}
\begin{document}\cmsNoteHeader{EXO-24-025}

\cmsNoteHeader{EXO-24-025}
\title{
Search for exotic Higgs boson decays \texorpdfstring{$\haa$}{H to AA} with \texorpdfstring{$\agg$}{A to gamma gamma} in events with a semi-merged topology in proton-proton collisions at \texorpdfstring{$\sqrt{s} = 13\TeV$}{sqrt(s) = 13 TeV}
}

\date{\today}

\renewcommand{\Pa}{\ensuremath{\mathcal{A}}\xspace}
\newcommand{\agg}{\ensuremath{\Pa\to\PGg\PGg}\xspace}
\newcommand{\hgg}{\ensuremath{\PH\to\PGg\PGg}\xspace}
\newcommand{\haa}{\ensuremath{\PH\to \Pa\Pa}\xspace}
\newcommand{\hag}{\ensuremath{\PH\to \Pa\Pa \to 4\PGg}\xspace}
\newcommand{\Br}{\ensuremath{\mathcal{B}}\xspace}
\newcommand{\rmol}{\ensuremath{r_{\text{M}}}\xspace}
\newcommand{\ma}{\ensuremath{m_{\Pa}}\xspace}
\newcommand{\mG}{\ensuremath{m_{\Gamma}}\xspace}
\newcommand{\Pepem}{\ensuremath{\Pep\Pem}\xspace}
\newcommand{\mggg}{\ensuremath{m_{\Gamma \Gamma \Gamma}}\xspace}
\newcommand{\mma}{\ensuremath{2\mathrm{D}\,m_{\Pa}}\xspace}
\mathchardef\mhyph="2D
\newcommand{\xsh}{\ensuremath{\sigma(\Pp\Pp \to \PH)}\xspace}
\newcommand{\pb}{\ensuremath{\unit{pb}}\xspace}
\newcommand{\mh}{\ensuremath{m_{\PH}}\xspace}
\newcommand{\masr}{\ensuremath{\ma\mhyph\text{SR}}\xspace}
\newcommand{\mhsb}{\ensuremath{\mh\mhyph\text{SB}}\xspace}
\newcommand{\mhsr}{\ensuremath{\mh\mhyph\text{SR}}\xspace}
\newcommand{\masb}{\ensuremath{\ma\mhyph\text{SB}}\xspace}
\newcommand{\mhsblo}{\ensuremath{\mh\mhyph\text{SB}_{\text{low}}}\xspace}
\newcommand{\mhsbhi}{\ensuremath{\mh\mhyph\text{SB}_{\text{high}}}\xspace}
\newcommand{\mhsblohi}{\ensuremath{\mh\mhyph\text{SB}_{\text{low}\,\text{(high)}}}\xspace}
\newcommand{\mhsrmasb}{\ensuremath{\mhsr\cap\masb}\xspace}
\newcommand{\mhsrmasr}{\ensuremath{\mhsr\cap\masr}\xspace}
\newcommand{\mhsbmasb}{\ensuremath{\mhsb\cap\masb}\xspace}
\newcommand{\mhsbmasr}{\ensuremath{\mhsb\cap\masr}\xspace}
\newcommand{\nsg}{\ensuremath{N(\ma|\mhsr \cap \masr)}\xspace}
\newcommand{\intlumi}{\ensuremath{\int\!\mathcal{L}\,\text{dt}}\xspace}
\newcommand{\pmhsblo}{\ensuremath{p(\ma|\mhsblo)}\xspace}
\newcommand{\pmhsbhi}{\ensuremath{p(\ma|\mhsbhi)}\xspace}
\newcommand{\pmhsb}{\ensuremath{p(\ma|\mhsb)}\xspace}
\newcommand{\pmhsr}{\ensuremath{p(\ma|\mhsr)}\xspace}
\newcommand{\fsb}{\ensuremath{f_{\text{SB}}}\xspace}
\newcommand{\fsblo}{\ensuremath{f_{\text{SB} \mhyph \text{low}}}\xspace}
\newcommand{\fsblonom}{\ensuremath{f_{\text{SB} \mhyph \text{low, nom}}}\xspace}
\newcommand{\fsbloalt}{\ensuremath{f_{\text{SB} \mhyph \text{low, alt}}}\xspace}
\newcommand{\fsbhi}{\ensuremath{f_{\text{SB} \mhyph \text{high}}}\xspace}
\newcommand{\nsblo}{\ensuremath{N_{\text{SB} \mhyph \text{low}}}\xspace}
\newcommand{\nsbhi}{\ensuremath{N_{\text{SB} \mhyph \text{high}}}\xspace}
\newcommand{\mam}{\ensuremath{m_{\Pa1}}\xspace}
\newcommand{\mar}{\ensuremath{m_{\Pa2}}\xspace}
\newcommand{\Itk}{\ensuremath{\mathcal{I}_{\text{tk}}}\xspace}
\newcommand{\zee}{\ensuremath{\PZ \to \Pep\Pem}\xspace}

\abstract{
A search for exotic Higgs boson decays \haa, with \agg is presented, using events with a semi-merged topology. One of the hypothetical particles, \Pa, is assumed to decay promptly into a semi-merged diphoton system reconstructed as a single photon-like object, while the other \Pa decays into two resolved photons. The search is performed using proton-proton collision data collected by the CMS experiment at $\sqrt{s} = 13\TeV$, corresponding to an integrated luminosity of 138\fbinv. The data agree with the standard model background expectation. Upper limits are set on the product of the Higgs boson production cross section and the branching fraction, $\xsh\,\Br(\hag)$, which range from 0.264 to 0.005\pb at 95\% confidence level, for \Pa masses in the range $1 < \ma < 15\GeV$. These limits are the most stringent to date in the 1--5\GeV \ma range.
}

\hypersetup{
pdfauthor={CMS Collaboration},
pdftitle={Search for exotic Higgs boson decays H to AA with A to gamma gamma in events with a semi-merged topology in proton-proton collisions at sqrt(s) = 13 TeV},
pdfsubject={CMS},
pdfkeywords={CMS, Higgs, beyond the standard model}}

\maketitle

\section{Introduction}
\label{sec:intro}

The standard model (SM) of particle physics continues to show
excellent agreement with measurements 
performed at the CERN LHC~\cite{LHCpaper} and
elsewhere. The search for physics beyond the SM~(BSM)
has led researchers to look beyond simple decay topologies and towards more
exotic signatures. Since the discovery of the Higgs boson (\PH) by the
ATLAS and CMS Collaborations~\cite{ATLAS:2012yve,CMS:2012qbp, CMS:higgslong2013} in 2012, the Higgs sector has remained a key area for BSM searches.
The latest CMS combination of Higgs boson measurements, which uses data from 2015--2018, limits the Higgs boson invisible (undetected) decay branching fraction to less than 13 (25)\% at 95\% confidence level (\CL)~\cite{CMS:2026combinedHiggsmeasurement}, leaving open the possibility of additional exotic Higgs boson decays. The Higgs sector is particularly well suited to probe BSM scenarios, as it may provide direct access to BSM states that do not participate in SM gauge interactions. The simplest extensions of the Higgs
sector include exotic decays of the form \haa, where \Pa is a hypothetical new
spin-0 particle decaying to a pair of SM particles~\cite{hexo}. Such
decays are found in BSM models containing an additional SM-neutral
singlet~\cite{hexo,hexo2}, such as the two-Higgs-doublet model plus singlet (2HDM+S), and are of interest in searches for axion-like
particle production~\cite{axiondmreview,pqtheory,axioncollider,axionreview}.

Searches for exotic Higgs boson decays of the type \haa have been performed at the LHC in several final states. 
For \Pa masses above the heavy-fermion thresholds $2m_{\Pgt}$ or $2m_{\PQb}$, the $\Pa \to \Pgt^+\Pgt^-$ or $\Pa \to \bbbar$ channels, respectively, are expected to become dominant, as predicted in certain 2HDM+S extensions~\cite{CMS:2024uru, CMS:2024zfv, CMS:2025hjt, ATLAS:2025qyn, CMSlonglivedjets}. Below these thresholds, the leptonic mode $\Pa \to\Pgmp\Pgmm$ becomes accessible and provides a clean experimental signature~\cite{CMS:2018mumu}. 
In contrast, the decay \agg, although loop-induced and generally subdominant, can become significant in photophilic scenarios where the \Pa couples to new uncolored charged particles, motivating searches for $\hag$~\cite{hexo}.

In such scenarios, reconstructing the diphoton invariant-mass spectrum becomes challenging for \Pa particles with masses of a few~\GeVns produced in \haa decays.
In the \Pa mass range of $\ma \lesssim 1\GeV$, \Pa has a large Lorentz
boost, causing the two photons from the \agg decay to be highly collimated. The separation between them becomes equal to or less than the Moli\`{e}re radius (\rmol = 2.2\unit{cm}) of
the material of the CMS crystal electromagnetic calorimeter (ECAL), which, in the barrel region, is comparable to the transverse crystal size. The two photons are predominantly reconstructed
as a single photon-like object by the standard CMS photon reconstruction algorithms~\cite{CMS:2017yfk}, as described in
Ref.~\cite{CMS:e2e-massreg}. For $\ma \gtrsim 1\GeV$, the separation between the two photons becomes more distinct. For example, for $1<\ma <2\GeV$, the two photons start to become distinguishable, with separations of around 3--5~\rmol at the ECAL surface. For $\ma \gtrsim 10\GeV$, the photons are well isolated and fully resolved, allowing the diphoton invariant mass to be reconstructed from the kinematics of both photon objects.

The CMS Collaboration reported on a search~\cite{CMS:Resolved} for \hag with four well-isolated photons in the final
state, with both photon pairs being fully
reconstructed, enabling a search in the \ma range of 15--62\GeV. The CMS Collaboration also performed a search in the highly boosted
regime, where each pair of photons is reconstructed as one merged photon-like
object, for \Pa masses in the 0.1--1.2\GeV range~\cite{CMS:Merged}.  
This direct search, the first of its kind in CMS, can directly measure the invariant mass of the merged diphoton candidates using a novel particle reconstruction technique that employs deep learning algorithms trained on the energy deposits in the ECAL~\cite{CMS:e2e-massreg}. The ATLAS Collaboration has also reported on an inclusive search for \hag decays over the full \ma range of 0.1--62\GeV~\cite{ATLAS:htoaato4g2023}.

The analysis presented in this paper focuses on events with three photon-like objects in the final state, to bridge the gap in the $1<\ma<15\GeV$ mass range between the two previous CMS searches mentioned above. The \Pa particle is assumed to decay promptly into two photons. 
In this so-called ``semi-merged'' mass regime, one of the \agg decays produces two well-separated and fully reconstructed photon objects, while the pair of photons from the other \Pa boson decay is reconstructed as a single photon object. This could occur either because of the actual merging of the photons, or detector and/or reconstruction energy thresholds. 
The merging of the photons is more common at lower masses, where the opening angle between the two photons is small, and their showers overlap within the ECAL. 
At higher \ma values, the photons are more separated, but the secondary photon is often too low in energy to be reconstructed because of the transverse momentum $\pt > 10\GeV$ threshold applied in the standard photon reconstruction~\cite{CMS:2020uim}.
A direct measurement of the invariant mass spectrum in this region by use of traditional methods is thus not possible.
Therefore, a novel end-to-end deep learning technique, called the ``mass regression'', was developed to directly reconstruct \ma from the pattern of energy deposits in the ECAL crystals.

The paper is organized as follows. 
Section~\ref{sec:Detector} provides a brief overview of the CMS detector. 
Section~\ref{sec:Samples} outlines the data and simulated samples used in the analysis. The event selection and object identification criteria are detailed in Section~\ref{sec:EventReco}. 
The mass regression technique based on a deep learning approach is presented in Section~\ref{sec:massreg}. The signal and background modeling are discussed in Section~\ref{sec:SignalBackground}. Section~\ref{sec:Systematics} summarizes the sources of systematic uncertainty.
Finally, Section~\ref{sec:Results} presents the results and their statistical interpretation, and Section~\ref{sec:Summary} provides a brief summary.
Tabulated results are provided in the HEPData record for this analysis~\cite{hepdata}.

\section{The CMS detector}
\label{sec:Detector}

The CMS apparatus~\cite{CMS:2008xjf,CMS:2023gfb} is a multipurpose, nearly hermetic detector, designed to trigger on~\cite{CMS:2020cmk,CMS:2016ngn,CMS:2024aqx} and identify electrons, muons, photons, and (charged and neutral) hadrons~\cite{CMS:2020uim,CMS:2018rym,CMS:2014pgm}. The central feature of the CMS apparatus is a superconducting solenoid
of 6\unit{m} internal diameter, providing a magnetic field of
3.8\unit{T}. Within the solenoid volume are a silicon pixel and strip
tracker, a lead tungstate crystal ECAL, and a brass and scintillator
hadron calorimeter (HCAL), each composed of a barrel and two endcap
sections. Forward calorimeters extend the pseudorapidity ($\eta$)
coverage provided by the barrel and endcap detectors. Muons are detected
in gas-ionization chambers embedded in the steel flux-return yoke
outside the solenoid.
A global ``particle-flow'' (PF) algorithm~\cite{CMS:2017yfk} aims to reconstruct all individual particles in an event by combining information from the all-silicon inner tracker, the ECAL and HCAL, and the muon system. The reconstructed particles are used to build \PGt leptons, jets, and missing transverse momentum~\cite{CMS:2018jrd,CMS:2016lmd,CMS:2019ctu}. Events of interest are selected using a two-tiered trigger system. The first level, composed of custom hardware processors, uses information from the calorimeters and muon detectors to select events at a rate of around 100\unit{kHz} within a fixed latency of 4\mus~\cite{CMS:2020cmk}. The second level, known as the high-level trigger, consists of a farm of processors running a version of the full event reconstruction software optimized for fast processing, and reduces the event rate to a few kHz before data storage~\cite{CMS:2016ngn,CMS:2024aqx}.

The ECAL consists of 75\,848 lead tungstate crystals, which provide a
coverage of $\abs{\eta} < 1.48 $ in the barrel region (EB) and
$1.48 < \abs{\eta} < 3.00$ in the two endcap regions. The full azimuthal
range $\abs{\phi} <\pi$ is covered for all regions.
In the EB, which is the focus of this paper, an
energy resolution of about 1\% is achieved for photons in the tens of
\GeVns energy range, for those that do not convert to an $\Pepem$ pair
before reaching the EB. The energy resolution for photon conversions in
the EB is about 1.6\% up to $\abs{\eta} = 1$, rising to about 3\% at
$\abs{\eta} = 1.48$~\cite{CMS:2023gfb}. Each EB crystal has
dimensions of $2.18\times2.18\times23\,\cm^3$. This corresponds to a transverse granularity of
$\Delta\eta \times \Delta\varphi = 0.0174 \times 0.0174$ for electromagnetic showers from particles originating in the center of the CMS detector. Since the crystal Moli\`{e}re radius is comparable to the crystal width, more than 90\% of the energy of photons
interacting only in the ECAL is laterally contained within a
$3{\times}3$ crystal matrix~\cite{CMS:2020uim}.

The material structure of the inner tracking detectors changes with
pseudorapidity. As a result, the number of radiation lengths ($X_0$)
that a photon traverses before reaching the EB surface varies from
$0.4\,X_0$ along the central plane ($\eta \approx 0$) up to $1.8\,X_0$
on the forward edge ($\eta \approx 1.4$). Correspondingly, the amount of
energy that a photon loses, and its probability to convert into an
$\Pepem$ pair before reaching the ECAL crystals varies with $\eta$.

Finally, an important feature of the proton collisions at the LHC is the
presence of additional, soft collisions concurrent with the primary
hard-scattering event. Such additional proton-proton ($\Pp\Pp$) interactions within
the same or nearby bunch crossings, known as pileup, can contribute
additional energy depositions, particularly in the ECAL. Since these energy deposits
may potentially be misidentified as photons, appropriate pileup
mitigation techniques are applied~\cite{puppi}.
A more detailed description of the CMS detector, together with a
definition of the coordinate system used and the relevant kinematic
variables, is reported in Refs.~\cite{CMS:2008xjf,CMS:2023gfb}.

\section{Data samples and simulated events}
\label{sec:Samples}

The analysis is performed using $\Pp\Pp$ collision data
at $\sqrt{s}=13\TeV$, collected with the CMS detector at the LHC from 2016
to 2018, corresponding to an integrated luminosity of 138\fbinv.
Simulated \hag samples generated with \ma = 1, 3, 5, 10, and 15\GeV
are used to define the signal model, as discussed in Section~\ref{sec:SignalBackground}. These values are chosen to evenly cover the signal mass phase space targeted in this analysis.
They are generated with \MGvATNLO(v2.5.5)~\cite{Madgraph:2014hca} at leading-order (LO) using the SM + Dark Vector + Dark Higgs boson model~\cite{hexo, hexo2} and a Higgs boson mass of 125\GeV with its SM width, with up to one associated jet and negligible lifetime of \Pa. While the underlying model can accommodate different lifetimes of \Pa depending on the couplings, only the prompt scenario is considered in this analysis. The parton distribution functions provided by NNPDF3.1~\cite{NNPDF:2017mvq} at next-to-next-to-leading order accuracy are used. The generated hard process is interfaced with \PYTHIA(v8.240)~\cite{Pythia8p2}
to simulate hadronization and the $\Pp\Pp$ underlying event, following the CP5~\cite{CP5Tune:2019csb} tune. The generated samples are processed through the CMS detector simulation based on \GEANTfour~\cite{GEANT4:2002zbu}, using the same reconstruction algorithms used on data. 
The dominant backgrounds in this search are SM events composed uniquely of jets produced through the strong interaction, referred to as quantum chromodynamics (QCD) multijet events, as well as $\PGg+\text{jet}$ and prompt diphoton production. The background contributions are modeled entirely from data in this analysis, and no simulated events are used to describe them.

Additional simulated minimum bias interactions are overlaid on all simulated events to account for pileup effects. Along with the simulated samples, interpolated signal shapes are also used to ensure comprehensive coverage.

\section{Event selection and reconstruction}
\label{sec:EventReco}

The events in the data are required to pass the diphoton high-level trigger selection used in the CMS low-mass \hgg search~\cite{CMS:2024yhz}, which requires the triggered diphoton objects to have an invariant mass greater than 55\GeV, with the photon minimum \pt thresholds of 30 and 18\GeV on the highest and second-highest \pt photons, respectively, as well as various calorimeter-based identification criteria.
Photon reconstruction~\cite{CMS:2020uim} in CMS starts by clustering ECAL crystal energy deposits. For photons with transverse energy greater than 10\GeV, clusters are grouped into superclusters to account for shower spread from tracker interactions and the magnetic field. Below this threshold, or if quality criteria fail, individual clusters may still be reconstructed as photon-like PF~\cite{CMS:2017yfk, ATLASPF} candidates.
Events that pass the trigger are required to pass the photon
reconstruction and kinematic selection requirements similar to those
used in the CMS SM \hgg search~\cite{hgg}.
Each photon candidate is required to satisfy standard CMS identification criteria that ensure the electromagnetic shower is compact, well-contained in the ECAL, and largely unaccompanied by hadronic energy or nearby tracks. These selections are designed to identify prompt, isolated photons and to suppress contamination from jets and electrons. The leading and subleading photons must have $\pt> 30$ and 18\GeV, respectively, and satisfy loose isolation and electron veto requirements~\cite{CMS:EGM-14-001}.

To account for the distinct kinematic and topological features of the \agg decay, the photon identification is further refined, following the same strategy as the CMS fully merged diphoton analysis~\cite{CMS:Merged}, as described below.
Discrimination between signal photons and background jets is achieved by using the photon vs.~electromagnetic (EM) -enriched jet multivariate discriminator~\cite{hgg}.
Here, EM-enriched jets refers to hadronic jets with a large EM energy fraction, typically originating from neutral meson decays such as $\Pgpz, \Pgh \to \PGg\PGg$.
As the opening angle between photons from \agg increases with $m_{\Pa}$, their showers become more similar to those produced by EM-enriched jets.
To maintain high signal efficiency in this regime, a looser requirement is applied to the discriminator output than in the CMS \hgg search. 
This relaxation naturally increases the rate of hadronic jet contamination, which is mitigated by applying a more restrictive requirement on the charged-hadron isolation, $I_{\text{ch}}$, defined as the transverse momentum of charged particles in a cone around the selected photon.
Selected events are required to contain exactly three reconstructed photons ($\Gamma$) that pass the above photon identification criteria, and have the invariant mass of the triphoton system within the range $90 < \mggg < 180\GeV$, which contains the Higgs boson signal mass window and sidebands.
This analysis only uses photons in the EB ($\abs{\eta} < 1.44$). This requirement retains approximately 80-90\% of the signal for $\ma$ ranging from 1 to 15\GeV, indicating that the majority of signal photons are found in the barrel. The use of the endcaps would require the development of a special strategy for the merged candidate mass regression in a complicated geometric environment. Studies allowing resolved photons in the endcaps, while requiring the merged candidate to be in the EB, show only a percent-level increase in the signal efficiency across the mass range considered. Therefore, the EB-only selection is adopted to maintain a uniform reconstruction strategy.
After applying all of the selection criteria to the simulated samples, signal events are selected with an estimated efficiency of 2.8--5.6\%. The overall signal efficiency is mainly limited by the trigger requirements and by the requirement of exactly three reconstructed photons in the event, which defines the semi-merged topology targeted in this analysis. At small \ma, where the fully merged topology becomes more prevalent, this requirement leads to a larger reduction in acceptance. Further reduction in efficiency arises from the photon identification and isolation requirements that affect all signal masses.

The three reconstructed photons that pass the
preselections are grouped into two classes of \Pa-particle candidates: merged ($\Pa1$) and resolved ($\Pa2$) candidates. The
resolved candidates $\Pa2$ are reconstructed by choosing the two closest
photons and using their momenta to obtain the invariant mass. From simulation, this is seen to correctly match the photons from the same parent \Pa particle 97--99\% of the time for various \ma. 
The merged candidate $\Pa1$ in semi-merged events is obtained from the third reconstructed photon, for which a direct measurement of the invariant mass is not possible with the reconstructed objects, and so the mass is reconstructed from the machine learning (ML) based regression.  

Each reconstructed \Pa-particle candidate is used to fill a two-dimensional (2D) \ma
shape template, which acts as either a signal or background
model. Under the identical mass hypothesis, $\mam = \mar$, the nominal signal would fall on the diagonal of this 2D plane. In addition, we also define an orthogonal signal region (SR) within the
Higgs boson mass window on the triphoton mass \mggg, under the assumption that the photon-like objects come directly from Higgs boson decays, with no additional invisible particles.
The SR and sideband (SB) regions along the \mggg axis are: 
\begin{itemize}
    \item \mhsblo: $90 < \mggg < 100\GeV$,
    \item \mhsr: $100 < \mggg < 135\GeV$,
    \item \mhsbhi: $135 < \mggg < 180\GeV$,
\end{itemize}

while the SR and SB along the $\mma$ axis are:
\begin{itemize}
    \item \masb : $\Delta \ma > 2\GeV$ (off-diagonal),
    \item \masr : $\Delta \ma < 2\GeV$ (on-diagonal).
\end{itemize}

The boundaries of the regions above are optimized using simulation to maximize the signal efficiency in the SR and minimize the signal contamination in the SBs, while maintaining a sufficient number of events in the SBs to have a reasonable background estimate. 
The shape and yield of the nonresonant background in the final SR are derived using the data sidebands in the \mggg and \mma regions described above.

\section{Signal mass regression using a graph neural network}
\label{sec:massreg}

Several LHC studies have explored the ML-based reconstruction of Lorentz-boosted particle decays~\cite{Ju_2020, ATLAS:2018wis, Komiske:2018cqr},  most often relying on PF candidates. While these studies show improved results over the traditional cutoff-based algorithms, their effectiveness in dealing with shower merging remains limited. This motivates an end-to-end strategy that minimizes intermediate preprocessing, enabling ML models to train directly on minimally processed detector data represented as detector images or permutation-invariant graphs, to predict the particle properties~\cite{Ghosh:2018gyw, Uboldi_2022, Andrews_2020, Adams_2019, ExaTrkX:2020nyf}. In CMS, data are organized into various tiers according to the level of processing~\cite{Petrucciani_2015MINIAOD, Peruzzi_2020NANOAOD}. While compact, clustered formats are optimized for most analyses, studies like this one require access to detailed calorimeter information available only in minimally processed data, which enables the reconstruction of merged-photon candidates using the mass regression described below.

In the semi-merged regime of the \hag decay explored in this search, one pair of photons from the
\Pa decay is resolved and reconstructed as two separate photons. At the same time,
the other pair could be highly collimated and reconstructed as a single
photon, or the secondary photon in this pair, while identifiable as a separate object, could remain unreconstructed. This can happen if
the secondary photon is well separated from the primary photon, but is
too low in energy ($\pt < 10\GeV$) to be reconstructed by the standard CMS PF algorithm, or the opening angle is large so that the secondary photon escapes into the endcap or falls out of the search window. In simulation, we find that at $\ma = 1\GeV$, the merged leg is dominated by true overlaps of the two photon showers, comprising more than 90\% of events. For larger masses, the semi-merged configuration with a soft secondary photon is the dominant contribution, accounting for roughly 75--95 (65--85)\% of events for \ma values in the range 3--5 (10--15)\GeV, corresponding to cases where the secondary photon is contained within the regression window. To handle these cases, the mass regression mentioned above is used to determine the underlying \Pa mass directly from the crystal-level patterns in the ECAL, an approach similar to that developed in Ref.~\cite{CMS:e2e-massreg} and used in Ref.~\cite{CMS:Merged}. For larger \ma, a larger fraction of events have a secondary photon that falls outside the regression window, resulting in a single-photon-like topology and reduced signal efficiency in the final SR.

While the approach described in Ref.~\cite{CMS:e2e-massreg} employs a convolutional neural network (CNN), this analysis uses a more advanced graph neural network (GNN), which has been found to outperform the CNN.
The choice of a GNN was motivated by several factors. Graph convolution operations can naturally learn rotational symmetries present in the
energy deposit patterns in the calorimeter in the presence of soft secondary photon deposits by utilizing permutation invariance, which is a challenge for CNNs, thus removing one level of complexity from the problem. 
The use of a graph format for the input data also enables the use of only those crystals that have energy deposited in them. This results in a substantial reduction of the problem size relative to CNNs, where the data are represented by large, very sparse, 2D grids.

\subsection{Model architecture and input images}
The mass regression is built with a GNN, called a dynamic
reduction network~\cite{Gray:2020mcm}, with residual connections added. The inputs to the nodes of the graph are the energy deposit maps from the EB. A
$128{\times}128$ matrix of ECAL crystals around the most energetic crystal (seed) of the reconstructed merged photon candidate is chosen to create an image array. This corresponds to a window within an angular distance $\DR  = \sqrt{\smash[b]{\Delta\eta\,^2 + \Delta\phi\,^2}} \approx 1.11$ around the seed. Given that the opening angle between the photons from the \Pa increases with its mass $\ma$, the width of this window is chosen such that it captures most of the secondary photon's energy deposits. More than 90\% of events fall inside the window for $\ma < 5\GeV$, and about 60--80\% for higher masses up to 15\GeV.

\begin{figure}[thp!]
\centering{
\includegraphics[width=0.32\textwidth]{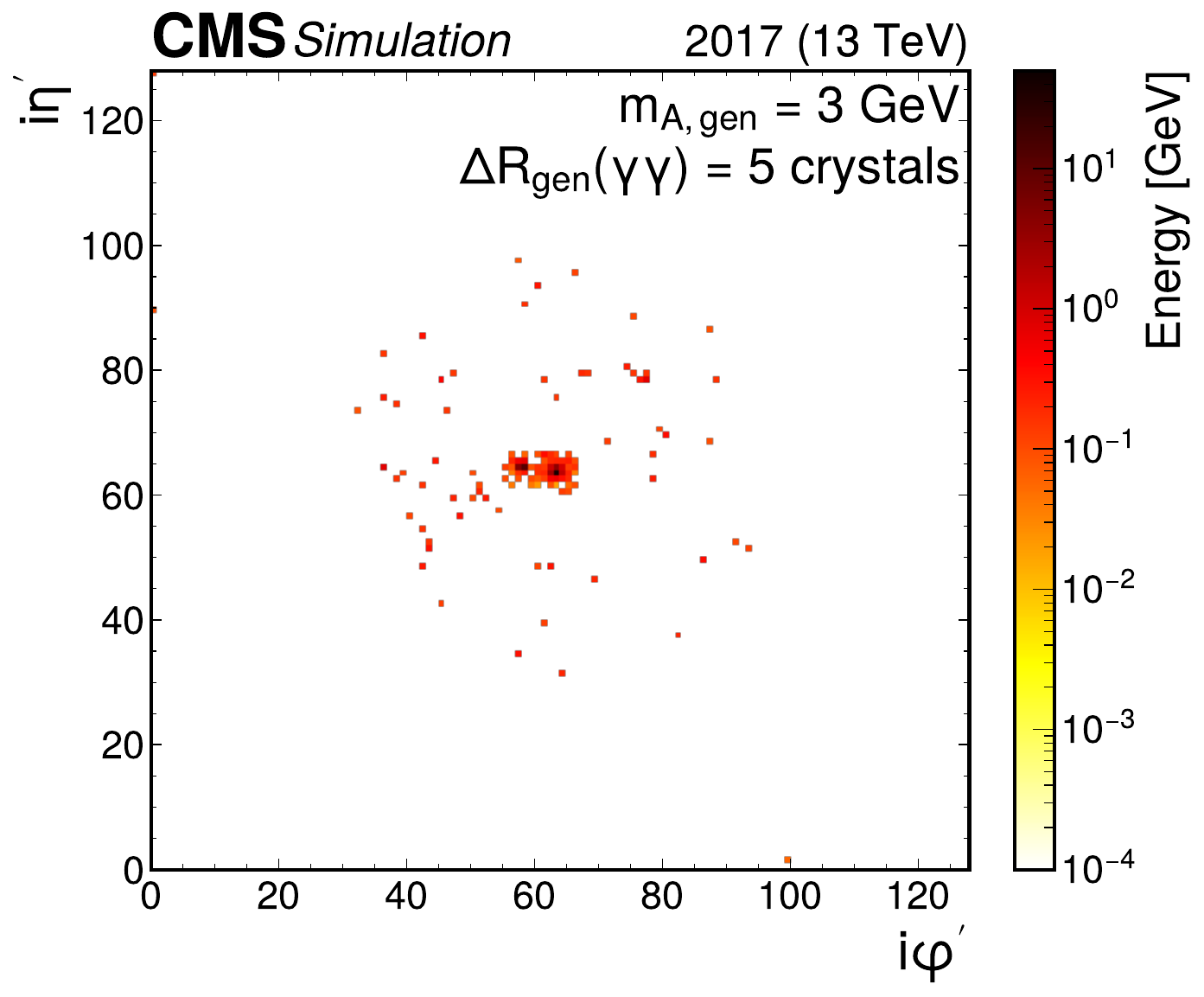} 
\includegraphics[width=0.32\textwidth]{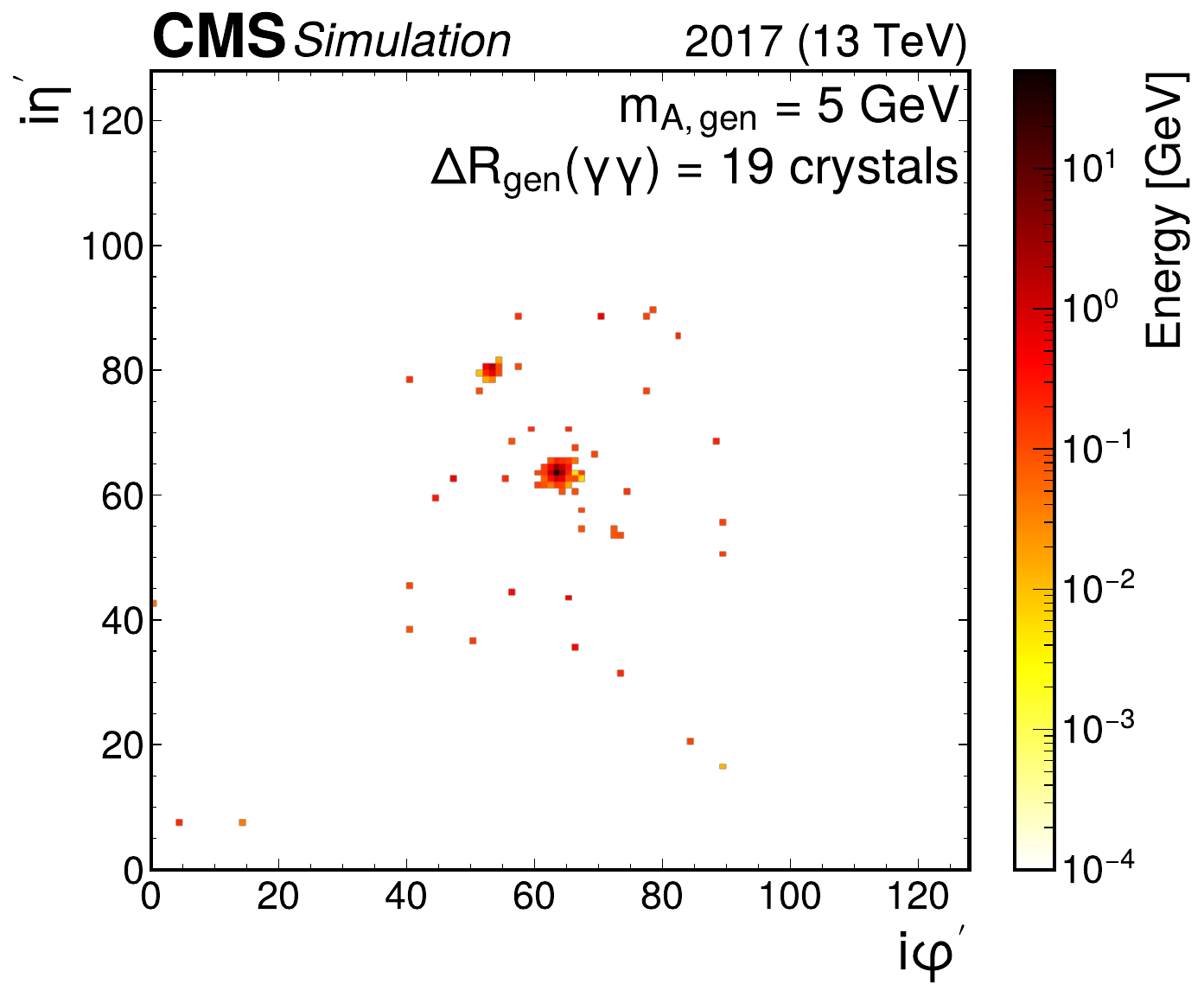} 
\includegraphics[width=0.32\textwidth]{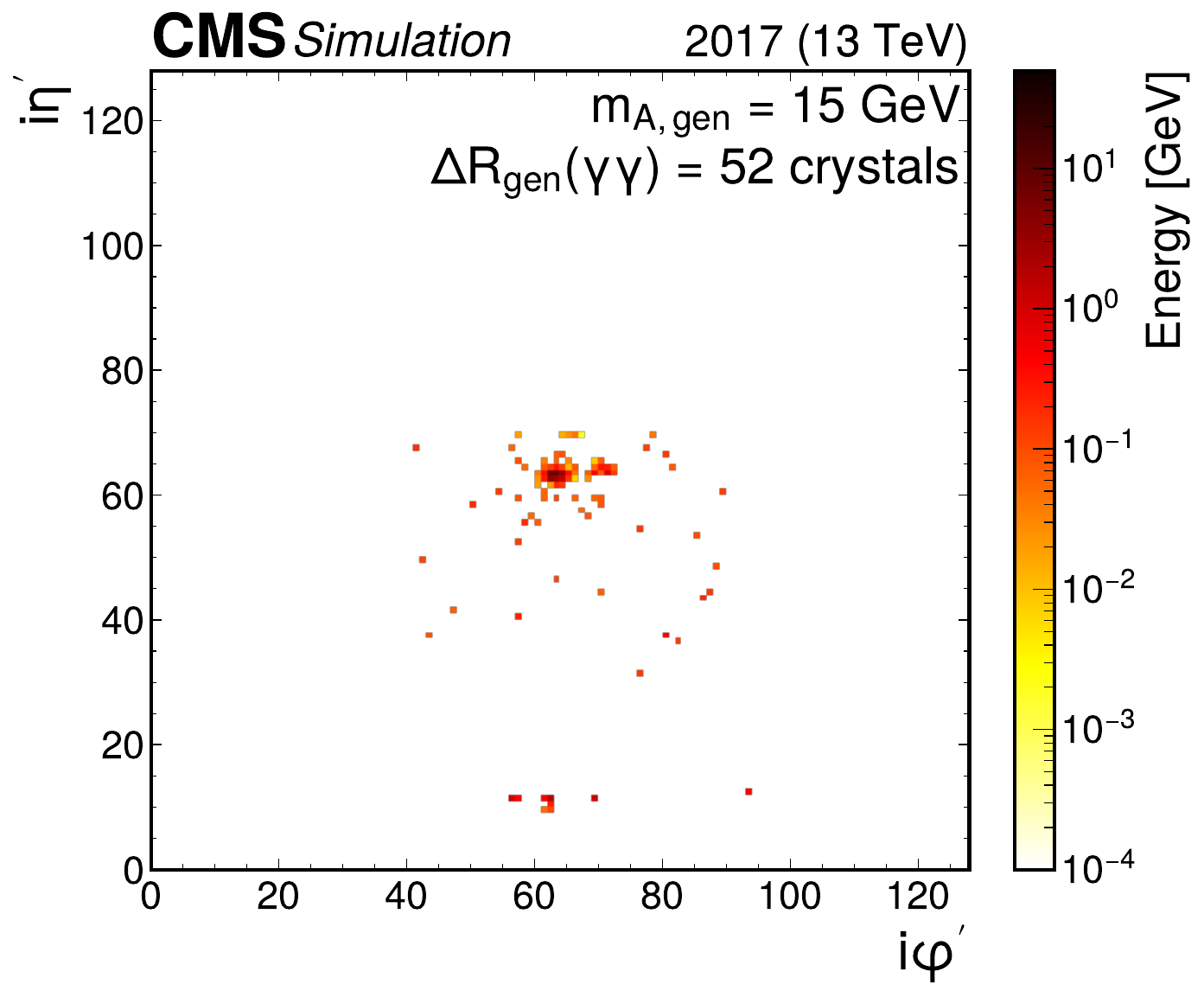} 
}
\caption{ Energy deposit maps in the ECAL from simulated \hag events corresponding to the 2017 data-taking conditions for the masses \ma = 3 (left), 5 (middle), and 15 (right) \GeV, shown in the relative ECAL crystal index coordinates ($i\eta^{'},\,i\phi^{'}$). Each image displays the ECAL deposits from a single \agg decay of the merged leg.
A close-up of the merged-photon candidate in the center of each image is shown, and the generator-level photon separation 
$\DR_{\text{gen}}(\PGg\PGg)$ is indicated above each panel for the corresponding mass point.}
\label{fig:inputimage}
\end{figure}

Examples of the ECAL energy deposit maps fed into the network are shown in Fig.~\ref{fig:inputimage}, taken from \hag simulation samples of different signal masses that correspond to different boosts and opening angles. The maps are plotted in the relative ECAL crystal index coordinates ($i\eta',\,i\phi'$), corresponding to the crystal positions along the $\eta$ and $\phi$ directions inside the cropped window around the merged-photon candidate.
Every pixel in the image directly represents the energy deposited in an individual ECAL crystal. These energy deposits correspond to the interaction between the incoming photon and the detector material. 

In Fig.~\ref{fig:inputimage}, the merged-photon candidate is shown at the center of each image, with the generator-level opening angle between the two photons, $\DR_{\text{gen}}(\PGg\PGg)$, printed above in terms of ECAL crystals. For the lowest mass point ($\ma = 3\GeV$, left), a second overlapping energy cluster is visible near the central deposit, consistent with two closely spaced photons being reconstructed as a single merged photon. At higher masses, the photons are more widely separated (as indicated by the larger opening angles); in the middle image, the secondary photon corresponds to the cluster to the upper left of the central deposit, while in the rightmost image, it corresponds to the low-energy cluster at the bottom. These secondary photon deposits are very low in energy and fall below the reconstruction threshold.

To fully utilize the differences in the EM shower development that result from the differences in the material budget in the $\eta$-direction in the detector, the ECAL energy deposit map is split into the transverse and longitudinal energy components, defined as  $\ET = E \sin{\theta}$ and $E_z = E \cos{\theta}$ respectively, where $E$ is the energy and $\theta$ is the polar angle of the energy deposit relative to the beam line. Additionally, the position coordinates of the crystals in $x$, $y$, and $z$ are also provided as separate features. These position coordinates, along with the relative location of energy deposits, aid in defining adjacency relations within the network and help the model to learn global variations in shower patterns due to differences in material budget as a function of $\eta$. 

\begin{figure}[bth!]
\centering{
\includegraphics[width=\textwidth]{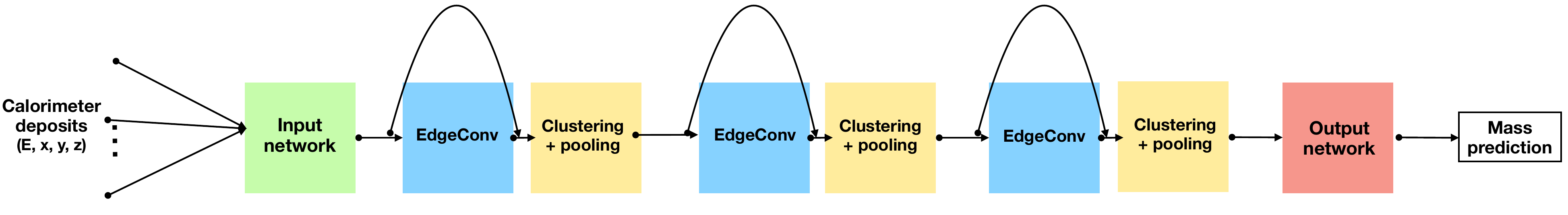} }
\caption{Illustration of the network architecture used for the mass regression. Input energy and position of the calorimeter deposits are processed by an MLP, followed by EdgeConv and clustering + pooling layers with residual connections. The final graph representation is passed through an output MLP to predict the object mass.}
\label{fig:networkarch}
\end{figure}

The architecture of the network is illustrated in Fig.~\ref{fig:networkarch}. The input nodes, which encode the information from the energy deposits, are first projected into a high-dimensional space using a multilayer perceptron (MLP)~\cite{Goodfellow-et-al-2016}, a fully connected neural network. The core of the model consists of edge convolution (EdgeConv)~\cite{wang2019dynamicgraphcnnlearning} modules. In each module, a graph is dynamically constructed by connecting each node to its nearest 16 neighbors in the learned feature space. The EdgeConv operation is then applied to these edges to update the node features. Following this, the original node features are added to those updated by the EdgeConv, forming a residual connection, before applying pooling. This residual connection, combined with pooling, progressively reduces the graph size. After the final clustering and pooling stage, global max pooling is applied across the remaining nodes, and the resulting feature vector is passed through an output MLP with a single node. This output is compared to the true mass of the object responsible for the input energy pattern using the mean absolute error as the loss function.

\subsection{Training}
\label{sec:mltraining}

For training, additional sets of simulated samples of \agg decays were produced with flat distributions in \ma and \pt, within the ranges
$\ma = (0,\,18)\GeV$ and $\pt = (10,\,200)\GeV$, and restricted to the EB, with $\abs{\eta} < 1.44$. 
The photon selection criteria defined for merged photon-like candidates preferentially select highly boosted \Pa particles, which tends to sculpt the $\pt^{\Pa}$ vs.~$\ma$ phase space.
To avoid biasing the network, the training samples are produced to ensure that the ($\pt^{\Pa}, \ma$) phase space is uniformly distributed after photon selection criteria have been applied.

At the boundaries of the training phase space, the performance of any deep neural network tends to degrade because those regions are underrepresented in the training data, leading to reduced generalization and biased predictions. To mitigate this effect and better equip the network to distinguish genuine merged-photon candidates from \Pa decays from single photons originating from background processes, simulated single-photon samples are also included in the training set. These single photons are assigned a random negative target mass value in the range $-3$ to 0\GeV, following the ``domain continuation'' method described in Ref.~\cite{CMS:e2e-massreg}.
The minimum value of $-3\GeV$ is optimized from training such that the mass response is linear, and with a minimal contribution of single photons in the mass domain of interest, $\ma > 0\GeV$.
By augmenting the training set with true photons not associated with the \agg decays and assigning a nonphysical negative mass value, we effectively reduce the background contamination by removing the samples that become regressed with masses below zero during inference, while making the full low-\ma region accessible to the network. 

\begin{figure}[bt]
\centering{
\includegraphics[width=0.45\textwidth]{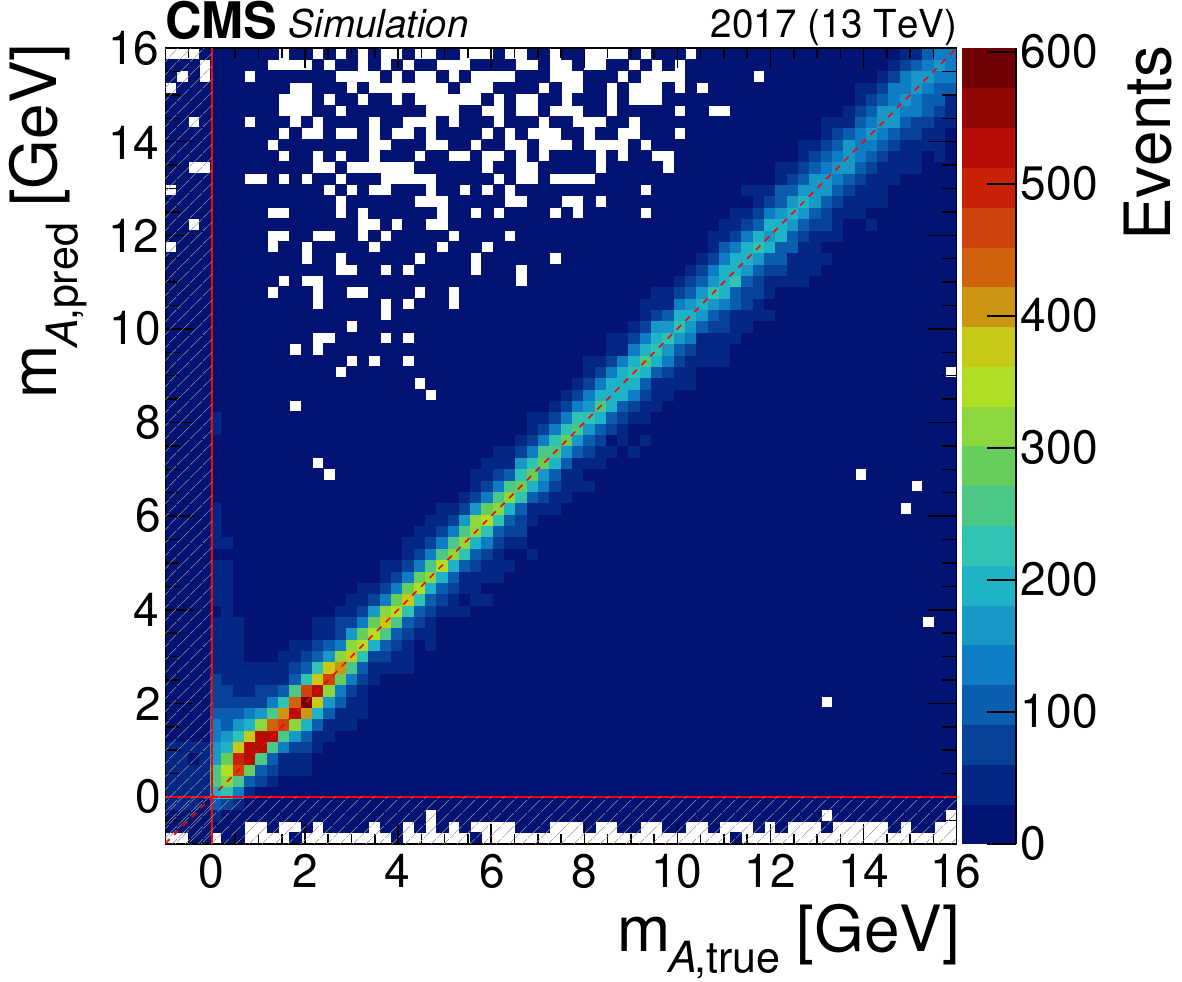} }
\caption{True vs.~predicted mass from the mass regression during training validation using simulated \agg samples corresponding to the 2017 data-taking conditions generated with a continuous uniform distribution in mass and \pt.
As discussed in Section~\ref{sec:mltraining}, the training includes domain continuation, which introduces negative target mass labels to improve linearity near the boundary; regions below and to the left of the solid red lines correspond to these negative mass predictions, which are not considered in the final selection.}
\label{fig:training}
\end{figure}

After the photon selection criteria are applied, the available samples, corresponding to the 2017 data-taking conditions, are split into independent training and validation subsets, with the latter used only to monitor performance during training. In total, about 600k \agg events and 70k single-photon events are used, split in an approximately 4:1 ratio between the training and validation sets.
The number of single-photon events is chosen such that the density of negative-mass samples is the same as that of the positive-mass samples. The regressed mass as a function of the true mass is shown in Fig.~\ref{fig:training}, demonstrating that the training works as intended: the regression learns the true mass efficiently over the entire mass range.

\subsection{Validation on simulation and data}
\label{sec:validation}

\begin{figure}[bt]
\centering{
\includegraphics[width=0.45\textwidth]{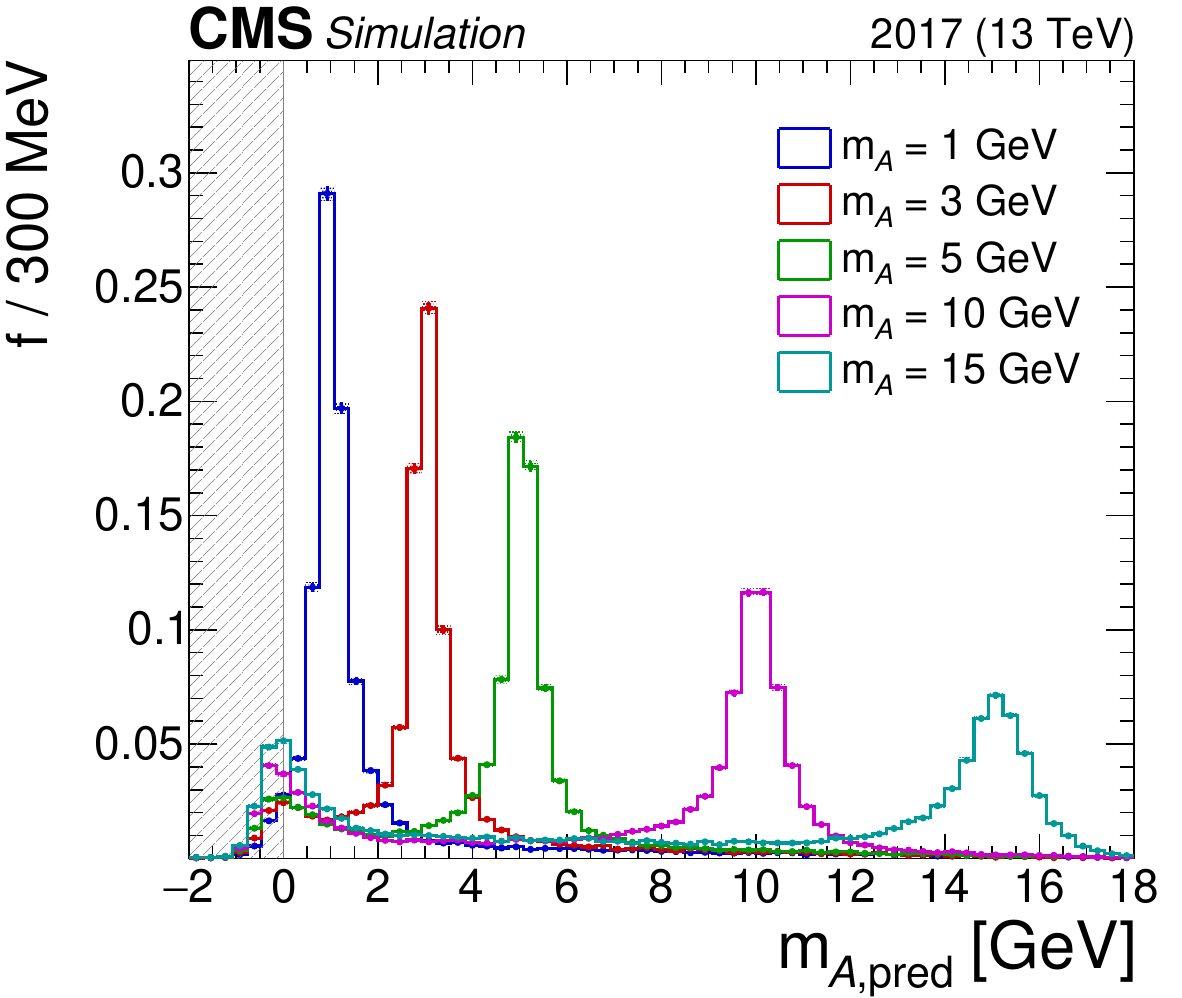} }
\caption{Predicted mass distribution from the mass regression on the merged leg of the simulated \hag events corresponding to the 2017 data-taking conditions, which pass the event selection criteria, for different simulated mass points. The variable $f$ denotes the fraction of events per bin. The area of each distribution is normalized to unity. Negative mass predictions shown in the hatched region are not considered for final selections. }
\label{fig:sig-overlay}
\end{figure}

\begin{figure}[!bthp]
\centering{
\includegraphics[width=0.44\textwidth]{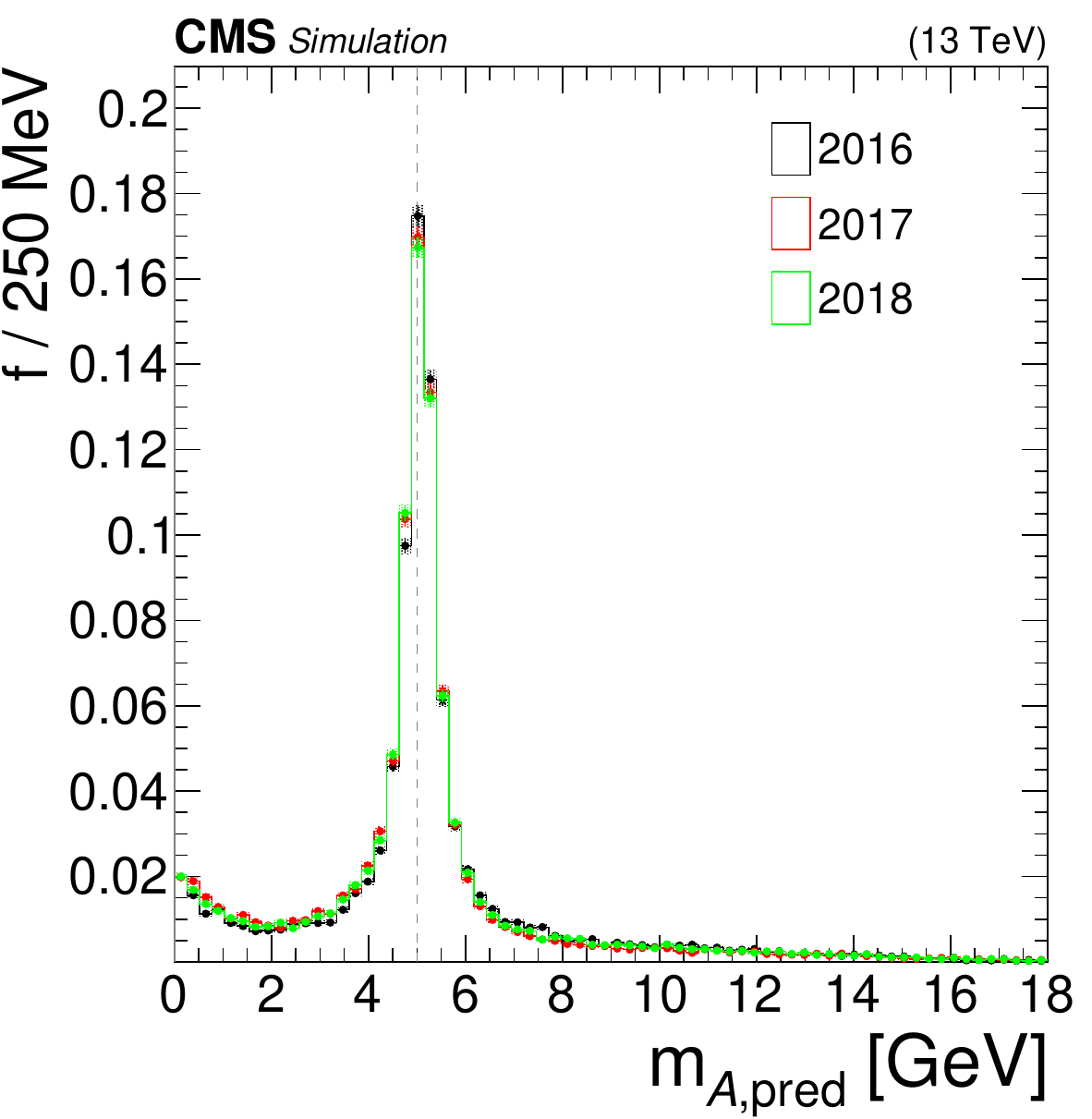} 
}
\caption{Predicted mass distribution from the mass regression on the merged leg of the simulated \hag events with $\ma = 5\GeV$, reconstructed under detector conditions corresponding to the 2016, 2017, and 2018 data-taking periods. The same regression and event selection are applied in all cases. The variable $f$ denotes the fraction of events per bin. The area of each distribution is normalized to unity.}
\label{fig:predmass}
\end{figure}

The regression model is validated on fixed-mass simulated samples of the signal \hag, for various \Pa particle masses. Figure~\ref{fig:sig-overlay} shows the mass regression prediction for the merged photon candidates that pass the selections, for the signal samples of various masses.
It can be seen that the regression can reconstruct the mass peaks reasonably well for all $\ma$ points. As the mass increases, there is an increasing peak at $\ma \sim 0\GeV$ because the ``merged leg'' looks more like a single photon, either because the secondary photon is extremely low in energy or the opening angle is too high, so it goes out of the image window. Figure~\ref{fig:predmass}, which shows simulated samples corresponding to different data-taking periods, demonstrates that the predicted mass peaks are well aligned for a signal mass of 5\GeV.
It highlights the robustness of the regression network to differences in conditions among different data-taking periods, given that it was trained exclusively using simulated samples corresponding to the 2017 data-taking period.

Since the signal models are built from simulated events and compared to background models derived from data, calibration and adjustment of the response of the mass regression between data and simulation are also performed.
Although decays of neutral mesons such as $\eta$ and \PGpz might seem suitable for this purpose, the selected events often contain highly collimated QCD jets with multiple mesons, which lead to shifted predicted masses. Additionally, limited simulation statistics and potential mismodeling of QCD processes make reliable comparisons with the data challenging. Another candidate, the \PJGy meson, was considered; however, selecting boosted \PJGy mesons that match the \Pa particle kinematics yielded an extremely low number of events, rendering it unsuitable for a meaningful calibration.

Instead, we turn to the decays of \zee, following the approach used in the fully merged diphoton search~\cite{CMS:Merged}. The electrons from the decay are produced in abundance and are well isolated from hadronic activity, similar to the case of the \agg decays. When interacting with the magnetic field of the CMS solenoid, the electron shower gets slightly smeared and can resemble a highly merged photon shower from the \Pa. The regression is validated on the \textit{single} electron showers which peak around $m \approx 0\GeV$. Electron candidates are obtained using the ``tag-and-probe'' method~\cite{CMS:WZpaper} and are required to be matched to a photon candidate passing the photon identification criteria. 
In addition to this, to better represent semi-merged-like events characterized by a soft secondary energy deposit near the primary photon deposit, we examine a specific subset of the selected events. These contain low-energy photon-like PF candidates mentioned in Section~\ref{sec:EventReco}. The \pt spectrum of these deposits has been verified in simulation to overlap with that of the secondary photon from \agg decays across all signal mass points, confirming the kinematic representativeness of this proxy.

\begin{figure}[!thp]
\centering{
\includegraphics[width=0.43\textwidth]{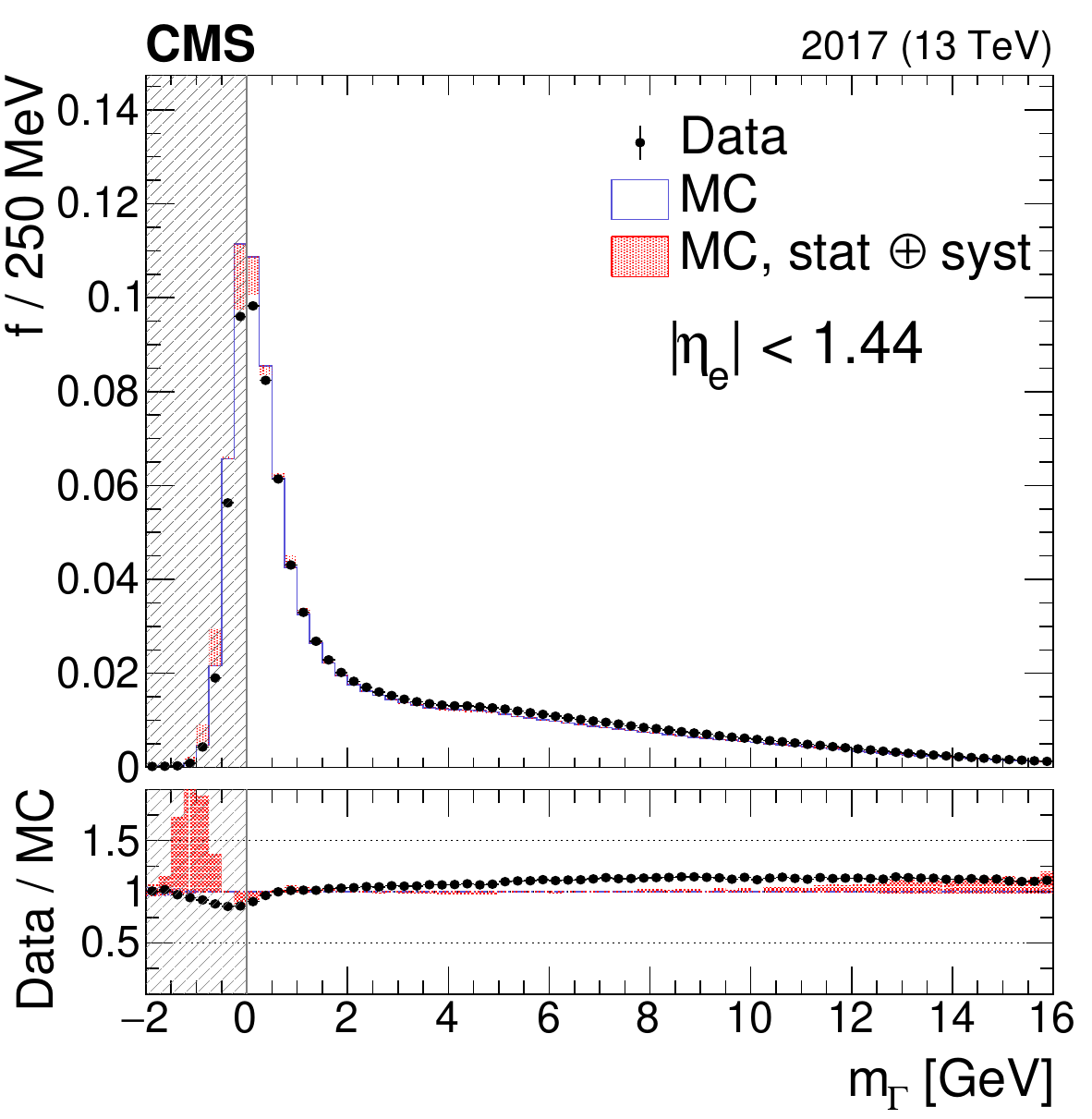} 
\includegraphics[width=0.43\textwidth]{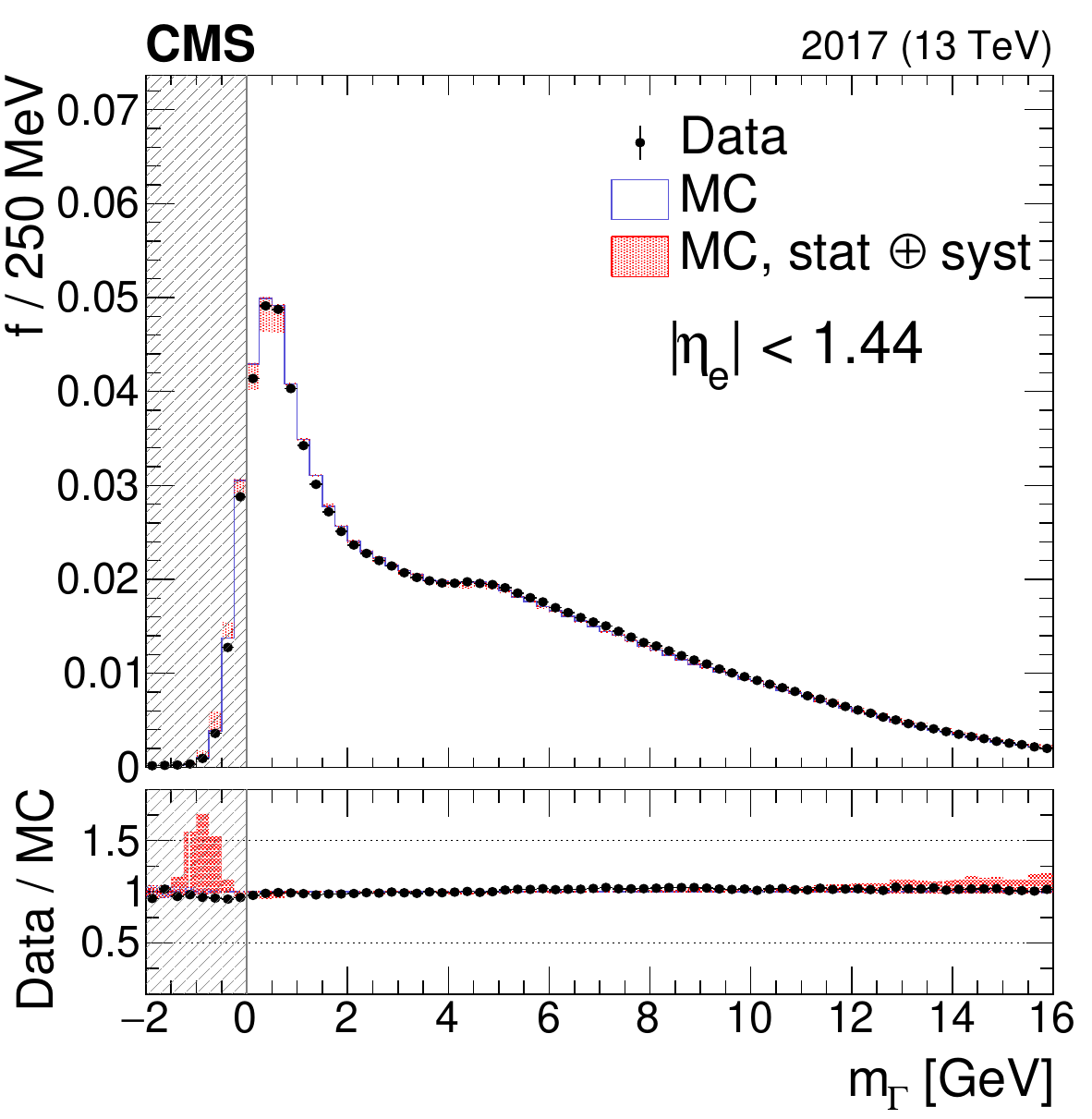}
}
\caption{Regressed mass for \zee electrons in the 2017 data vs.~simulation for all events passing the selections (left), showing a peak at zero, and for a subset of those events with a soft photon-like deposit near the electron (right), showing a peak away from zero. The variable $f$ denotes the fraction of events per bin. The area of each distribution is normalized to unity. In the upper panels of each plot, the coverage of the best fit scale and smearing estimates in simulated events (``MC, stat $\oplus$ syst''), plus statistical uncertainties added in quadrature, are plotted as the red bands around the original simulated sample, which are shown as the blue lines (``MC''). The data events are shown as black points, with statistical uncertainties indicated by the error bars. In the lower panels of each plot, the ratio of data to simulation is shown with the black points, where the error bars represent the statistical uncertainties in the data. The red-filled area represents the combined statistical and systematic uncertainty in the best fit corrected simulation, shown relative to the nominal prediction as an uncertainty envelope around unity. The gray hatched regions correspond to events that do not enter the selection.}
\label{fig:Zee-calibration}
\end{figure}

The resulting mass distributions are shown in Fig.~\ref{fig:Zee-calibration}: the left panel corresponds to all selected \zee events, peaking near zero, while the right panel shows the subset with a nearby soft PF deposit, producing a shifted distribution characteristic of semi-merged topologies. 
The relative response between data and simulation is parameterized using two components: a mass-scale variation, which adjusts the overall scale of the simulated regressed mass, and a smearing variation, which introduces extra Gaussian smearing to match the resolution observed in data.
These variations are evaluated separately for each data-taking year and for three EB pseudorapidity regions: central ($\abs{\eta} < 0.5$), middle ($0.5 < \abs{\eta} < 1.0$), and forward ($1.0 < \abs{\eta} < 1.44$). A representative example using 2017 data and simulation over the full EB pseudorapidity range is shown in the figure. The derived variations are incorporated as systematic uncertainties in the signal model, as discussed in Section~\ref{sec:Systematics}.
In Fig.~\ref{fig:Zee-calibration},
the shaded red band corresponds to the uncertainty envelope from the scale and smearing variations. 
These are applied as one-sided systematic variations (shift or no shift), and the resulting envelope spans the possible upward or downward shift of the prediction relative to the nominal simulation.
The data and simulation show agreement within the assigned uncertainties, with the residual differences at the percent level. This provides a validation of the calibration in configurations relevant for semi-merged topologies as well.

\section{Signal and background models}
\label{sec:SignalBackground}

\subsection{Signal model}
\label{sec:signalmodel}

The final SR is defined as \mhsrmasr.
In this region, we construct a signal model that specifies both the shape of the distribution and its total yield (normalization) 
with respect to the data for a given integrated luminosity, for each \ma hypothesis. 
The shape is derived by creating the \mma template using the simulated
\hag events passing the event selection criteria described in
Section~\ref{sec:EventReco}. 
Simulated samples are generated at nominal \Pa mass points of \ma = 1, 3, 5, 10, and 15\GeV, for which a shape template is produced and compared against the estimated background model.

\begin{figure}[!htbp]
\centering
\includegraphics[width=.49\textwidth]{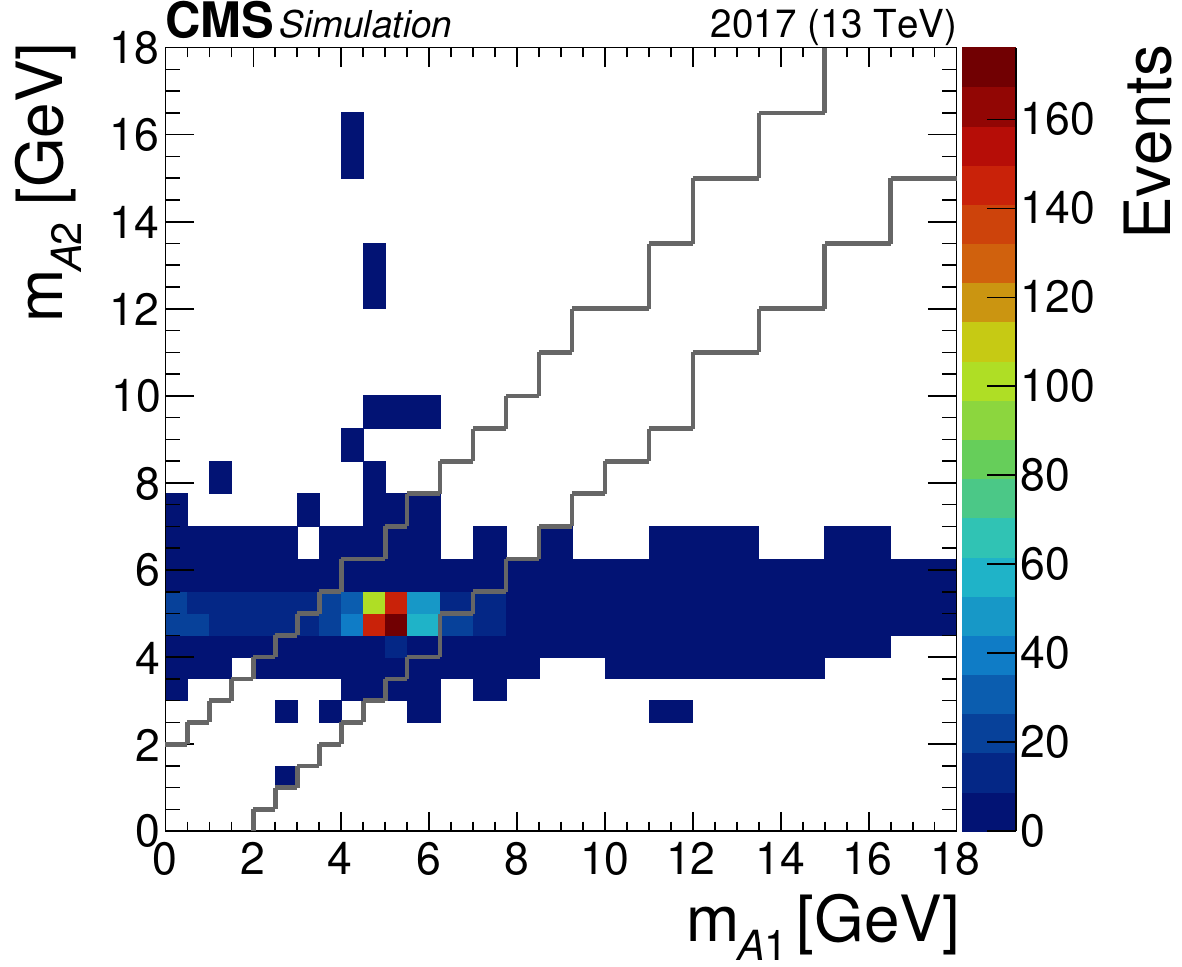} 
\caption{The \mma distribution of the signal model from 2017 simulation for the $\ma= 5\GeV$ hypothesis, normalized to $\xsh\Br (\hag) = 1\pb$. The region enclosed by the jagged lines defines the \masr region, while the area outside corresponds to the \masb region.}
\label{fig:sg-mA5Gev}
\end{figure}

The yield or normalization \nsg in the final SR for a particular \ma hypothesis with the data corresponding to a given integrated luminosity $\intlumi$ is given by:
\begin{equation}
    \nsg = \xsh\,\Br(\hag)\,\varepsilon(\ma|\mhsrmasr)\int\!\mathcal{L}\,\text{dt},
\end{equation}
where \xsh is the production cross section of the Higgs boson, $\Br (\hag)$ is the branching fraction of \hag, and  $\varepsilon(\ma|\mhsrmasr)$ is the efficiency of selection of the \hag events into the final SR, for a given \ma hypothesis. For ease of calculation, we choose a hypothetical signal cross section $\xsh\Br (\hag) = 1\pb$ to normalize the templates. If we consider the SM Higgs boson production cross section  $\xsh = 55.6\pb$~\cite{LHCHiggsCrossSectionWorkingGroup4:2016ypw}, this is equivalent to $\Br(\hag) = 1.8\times 10^{-2}$. 
Intermediate signal mass models are constructed by interpolating both the shape and normalization between the generated mass points with $\Delta \ma < 2\GeV$, ensuring smooth signal coverage across the full \ma mass range. We validate the interpolation method using generated signal mass points and find the shapes to be comparable.
Figure~\ref{fig:sg-mA5Gev} shows the \mma distribution of the signal model from simulation for the $\ma = 5\GeV$ hypothesis.
The region enclosed by the jagged lines defines the \masr region, while the area outside corresponds to the \masb region.
The \mma distributions are binned using variable bin widths, chosen to maintain sufficient statistical power in each bin. Within this binning scheme, the final template contains 119 bins, with the nominal signal lying along the diagonal within the \masr, centered around the expected value of 5\GeV for both axes.

\subsection{Background model}
\label{sec:bkgmodel}

The background estimation is performed using data in the SB regions. The shape and yield of the background model are derived by constructing a \mma template, using the data SBs in both the \mh and \mma distributions. 

To obtain the shape of the background template, we assume that the \mma shape in the \mhsr can be expressed as a linear sum of the corresponding shapes in the \mhsblo and \mhsbhi regions:
\begin{equation} \label{eq:bgshape}
    \pmhsr = \fsblo \pmhsblo + \fsbhi \pmhsbhi,
\end{equation}
where $p(\ma|\mh)$ denotes the corresponding \mma shape in each region, with
\begin{equation} \label{eq:bgfracsum}
    \fsblo + \fsbhi =  1.
\end{equation}
The $\fsblo\,(\fsbhi)$ is the relative contribution of the $\mhsblo\,(\mhsbhi)$ shape to the total shape of the \mhsb, defined as:
\begin{equation}
    \fsblo = \frac{\nsblo}{\nsblo + \nsbhi}.
\end{equation}
We then assume that the ratio of the number of events along the \mma diagonal region (\masr) derived from \mhsb to the number of events observed in the \mhsr is the same as that in the off-diagonal region (\masb): 
\begin{equation}
    \frac{N(\mhsrmasr)}{N(\mhsbmasr)} = \frac{N(\mhsrmasb)}{N(\mhsbmasb)}.
\end{equation}
Taking the ratio from the off-diagonal region, which is signal-depleted by construction, and the on-diagonal yield from the \mhsb, we can obtain a background model for the final SR:
\begin{equation} \label{eq:bgyield}
    N(\mhsrmasr) =  \frac{N(\mhsrmasb)}{N(\mhsbmasb)} N(\mhsbmasr).
\end{equation}
Thus, the final background model for the SR, combining the shape from Eq.~\eqref{eq:bgshape} and the yield from Eq.~\eqref{eq:bgyield}, is:
\begin{equation}
\label{eq:bgfinalbeforept}
\begin{split}
    \textrm{bkg}(\mhsrmasr) & = N(\mhsrmasr) \pmhsr \\
                           & =  N(\mhsbmasr) p(\ma | \mhsbmasr) \\
                           & \times \frac{N(\mhsrmasb)}{N(\mhsbmasb)}.
\end{split}
\end{equation}

\begin{figure}[!htbp]
\centering
\includegraphics[width=.49\textwidth]{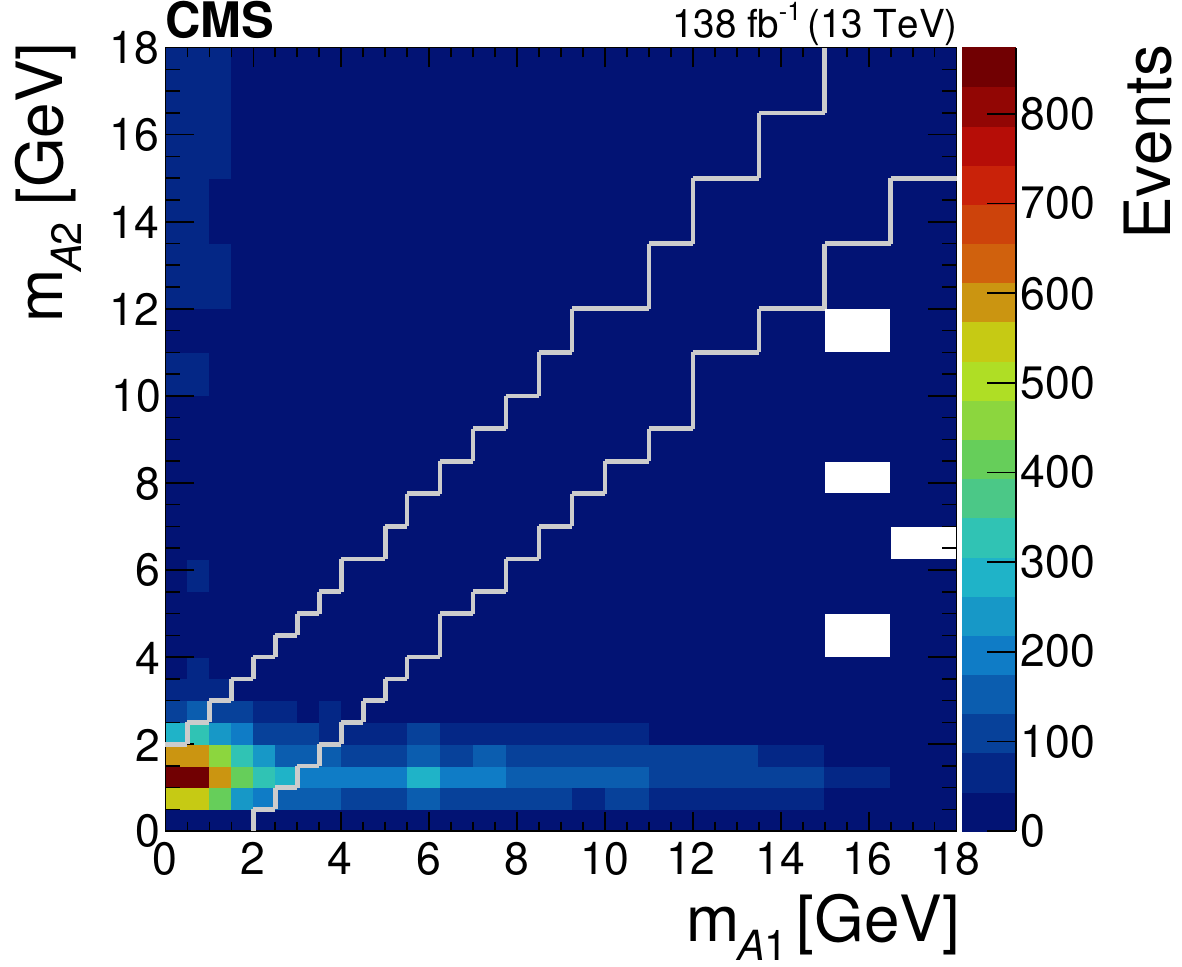} 
\caption{The \mma distribution of the background model constructed from data sidebands. The region enclosed by the jagged lines defines the \masr region, while the area outside corresponds to the \masb region.}
\label{fig:2d-bkg}
\end{figure}

To visualize the resulting background model, we construct the full \mma 
distribution using the procedure described above. This distribution, binned using the same variable binning scheme as the signal model, is shown in Fig.~\ref{fig:2d-bkg}. The region enclosed by the jagged lines defines the \masr region, while the area outside corresponds to the \masb region.

\subsubsection{Background model validation}
\label{sec:bkgval}

In order to validate the background estimation procedures described above, we compare the expected and observed \mma distribution in the \masb regions, where signal contamination was considered but deemed to be negligible. 
Any residual differences 
between the two distributions are quantified by fitting the bin-by-bin residuals in the 
1D projections along each mass axis with Chebyshev polynomials. The resulting functions are 
then applied to the nominal 2D background distribution to produce alternative templates, 
which are used to evaluate the background shape systematic uncertainties discussed in 
Section~\ref{sec:Systematics}.

\begin{figure}[!htbp]
\includegraphics[width=.99\textwidth]{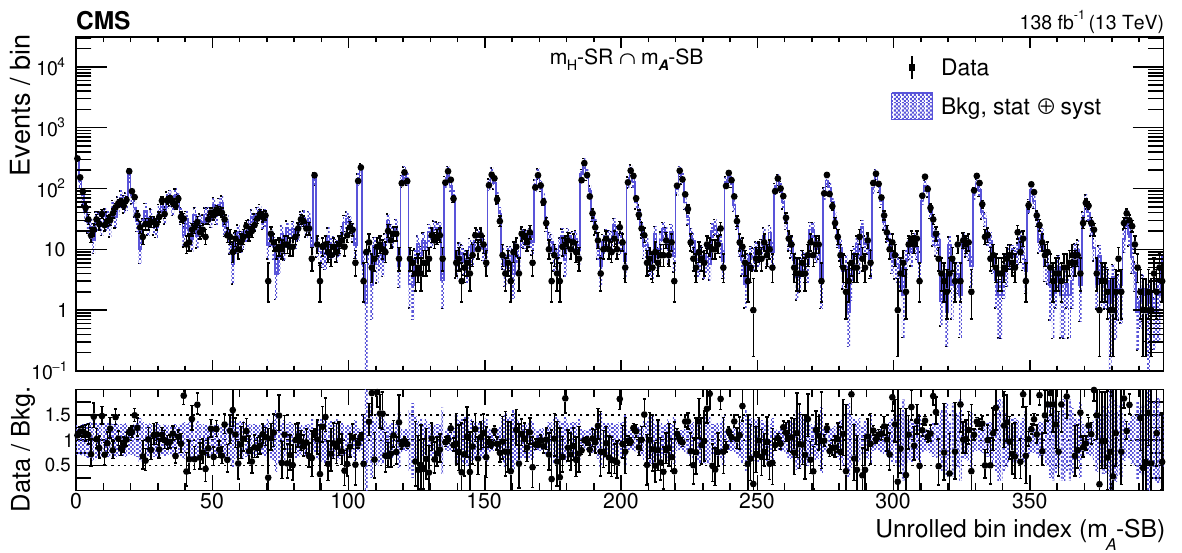}\\
\centering
\includegraphics[width=.49\textwidth]{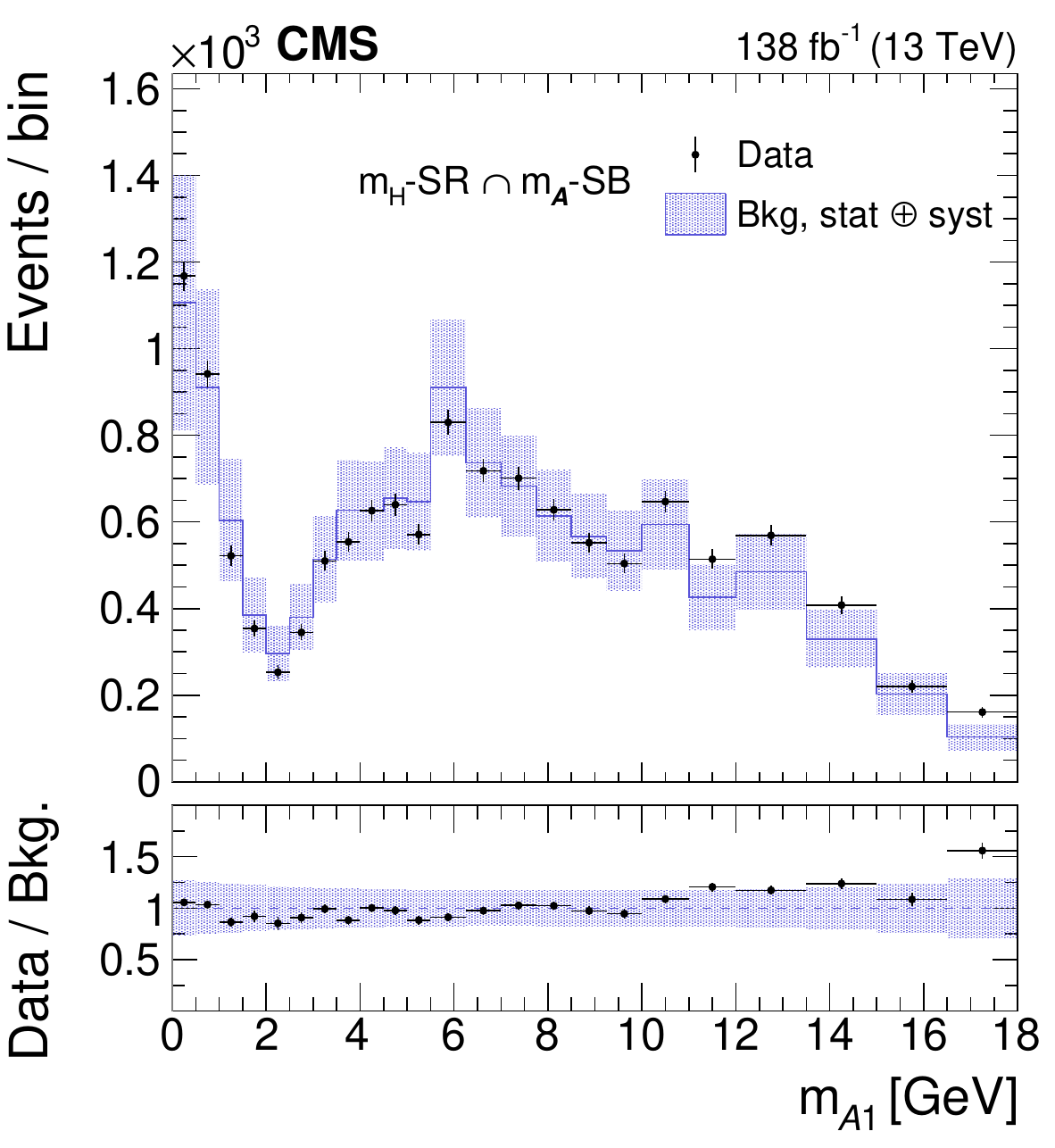}
\includegraphics[width=.49\textwidth]{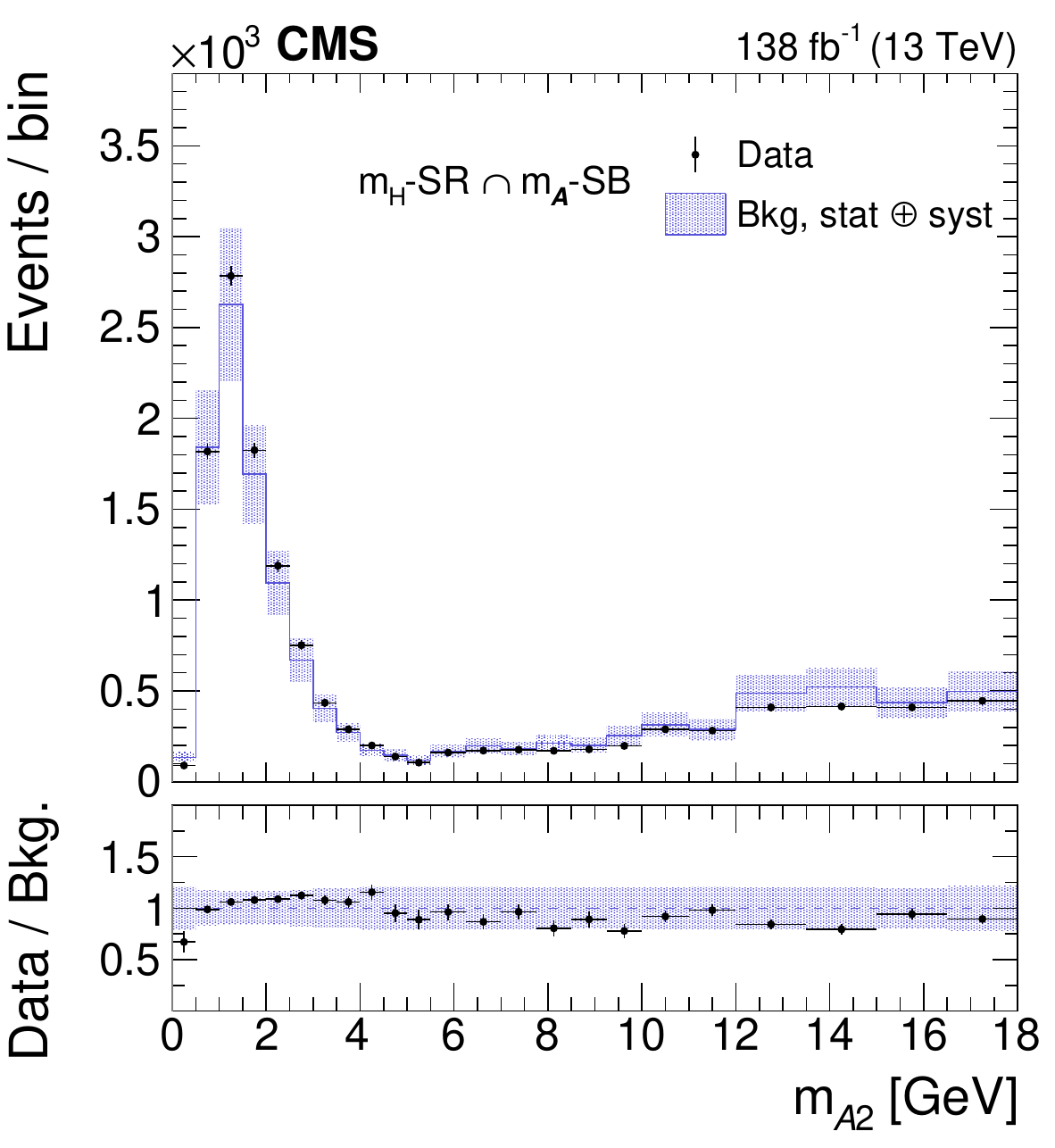}
\caption{Expected background versus observed data \mma spectra in the \mhsrmasb region. Upper plot: unrolled spectra made by scanning along bins of increasing \mar at fixed \mam before incrementing in \mam. Only the bins in the \masb region are included, with the x-axis corresponding to the unrolled bin index of the selected bins, listed sequentially. 
Lower plot: projected 1D \ma spectra corresponding to merged (left) and resolved (right) candidates. In the upper panels of each plot, the black points with error bars (``Data'') are the observed data values, with the error bars corresponding to statistical uncertainties. The blue line corresponds to the background model, with the blue band corresponding to its statistical uncertainties and systematic uncertainties added in quadrature (``Bkg, stat $\oplus$ syst'').  In the lower panel of the same plot, the ratio of the observed data over the estimated background value is shown as black points, with error bars corresponding to statistical uncertainties in the former. The ratio of the statistical plus systematic uncertainties in the background over the background prediction is shown as the blue band.
}
\label{fig:bg:nom-nom:mA-SB-shift}
\end{figure}

The unrolled distribution of the \mma distribution in the \masb region is
plotted in Fig.~\ref{fig:bg:nom-nom:mA-SB-shift}. Only bins that belong to the \masb region are retained. The unrolling begins at the lower-left corner of the 2D histogram and scans vertically in \mar at fixed \mam, then steps to the next \mam column. Bins outside the \masb region are skipped, with the x-axis indexing the retained bins in this scanning order.
The one-dimensional (1D) projections of the distributions along each \ma axis are also shown in the lower panels.
In the upper panel of the plots, the black points are the observed data with error bars corresponding to the statistical uncertainties. The blue line shows the nominal expected background distribution, with the blue shaded region indicating its statistical and systematic uncertainties added in quadrature. In the lower panel, the
ratio of the observed data to the nominal expected background is shown as black points, with the error bars corresponding to statistical uncertainties in the former. The ratio of the statistical plus systematic uncertainties in the background over the background model is shown as the blue band. We find good agreement between the data and the background prediction within the uncertainties. In addition, signal injection tests are performed, which do not show any potential bias in the background prediction method for the entire range of signal hypotheses considered. As a further cross-check, the normalization factors in the signal region, that are derived from the combined sidebands, are found to be compatible with those estimated from the low and high sidebands separately. This demonstrates that the \mma shape does not vary significantly across the \mggg range and residual correlations between \mggg and the \mma structure do not bias the background estimate.

\newpage
\section{Systematic uncertainties}
\label{sec:Systematics}
The dominant uncertainty that affects the extraction of the signal is the statistical uncertainty in the background modeling. In addition, systematic uncertainties related to both the background and signal models are taken into account.
For the signal model, alternate templates are generated by shifting the nominal prediction up or down by one standard deviation of each uncertainty source. The following sources are considered:

\begin{itemize}
    \item \textit{Integrated luminosity}: The uncertainty in the integrated luminosity measurement is 1.2, 2.3, and 2.5\% for the 2016, 2017, and 2018 data-taking periods, respectively~\cite{CMS:lumi2016, CMS-PAS-LUM-17-004, CMS-PAS-LUM-18-002}, leading to an overall uncertainty of 1.6\%. These uncertainties are considered to be partially correlated across the three years.
    \item \textit{Trigger efficiency}: To account for differences in trigger efficiency between data and simulation, trigger efficiencies in data are derived via the tag-and-probe method~\cite{CMS:WZpaper} using \zee events. The efficiency estimated in simulation is applied to the signal model, and a systematic uncertainty is assigned to account for potential differences with respect to data. This uncertainty ranges from 20\% at $\ma = 1\GeV$ to 1\% at $\ma = 15\GeV$. Larger values are assigned at low \ma, where the kinematic phase space of the photon candidates that pass the trigger requirements differs slightly from that used in the standard trigger efficiency measurements. A conservative uncertainty, corresponding to the difference between the trigger efficiency in simulation and that obtained from data measurements, is assigned to ensure full coverage.
    \item \textit{Photon identification}: Correction factors are applied to the signal model to account for differences in photon identification efficiency between data and simulation, as computed in Ref.~\cite{CMS:Merged}. These scale factors are obtained with the tag-and-probe procedure using \zee events, where the smeared electron shower serves as a proxy for merged-photon candidates, directly establishing the validity of this approach for the merged-diphoton topology. The associated uncertainty in this correction ranges from 1 to 3\%. 
    \item \textit{Photon energy scale and resolution}: Corrections are applied to the photon energy scale and resolution in simulation, as computed in Ref.~\cite{CMS:2020uim}. The resulting uncertainty in the signal model is a maximum bin-by-bin variation of $<$3\% from the scale and $\sim$0.1\% from the resolution. 
    \item \textit{Mass regression scale and smearing}:
   Following the procedure described in Section~\ref{sec:validation}, the difference between the mass regression response in data and simulation is accounted for in the signal model by applying a relative scale shift and an additional Gaussian smearing to the simulated regressed mass of each merged-photon candidate. The scale and smearing parameters are determined independently for each data-taking year and each $\abs{\eta}$ category.
   The best-fit values of the relative scale, $\hat{s}_{\text{scale}}$, and smearing, $\hat{s}_{\text{smear}}$, are obtained by minimizing the $\chi^2$ between the regression response for electrons in data and in simulation after applying the corrections to the simulated distribution. The resulting scale shifts range from 1.6 to 6\%, and the additional smearing reaches up to 260\MeV.
\end{itemize}

For the background model, a systematic uncertainty is assigned to account for potential 
mismodeling of the expected background shape. As described in Section~\ref{sec:bkgval}, the residual 
differences between the observed and expected distributions in the \masb region are fitted 
along each mass axis using Chebyshev polynomials: second order in \mam and third order in 
\mar. Each polynomial coefficient defines an independent variation, where the full correction 
associated with that coefficient is applied as an up or down shift relative to the nominal 
2D background template, thereby producing the alternative templates used as the background shape 
systematic variations in the final fit. The impact of these background modeling uncertainties on the limits is subdominant compared to the statistical uncertainty of the background template.

\section{Results}
\label{sec:Results}

A statistical hypothesis test using a maximum likelihood estimation~\cite{CLsMethod}
is performed to extract the best fit between the observed \mma distribution and the constructed signal-plus-background model, $\mu S + B$, where $\mu$ is the signal strength parameter for each signal mass point. 
Systematic and statistical uncertainties from the constructed signal and background models are treated as penalty terms or ``nuisance parameters'' in the likelihood function. 
The results of this analysis are used to set upper limits on the product of the production cross section of the Higgs boson and the branching fraction into four photons via a pair of \Pa particles, $\sigma(\PH)\Br(\hag)$. The CL$_{s}$ criterion~\cite{Junk-CL:1999kv, CLs:2002hq} is used to set the upper limits, using the asymptotic approximation~\cite{Cowan_2011}.
The results are determined using the CMS statistical analysis tool \textsc{Combine}~\cite{CMS:COMBINE}.

\begin{figure}[!htbp]
\centering
\includegraphics[width=.98\textwidth]{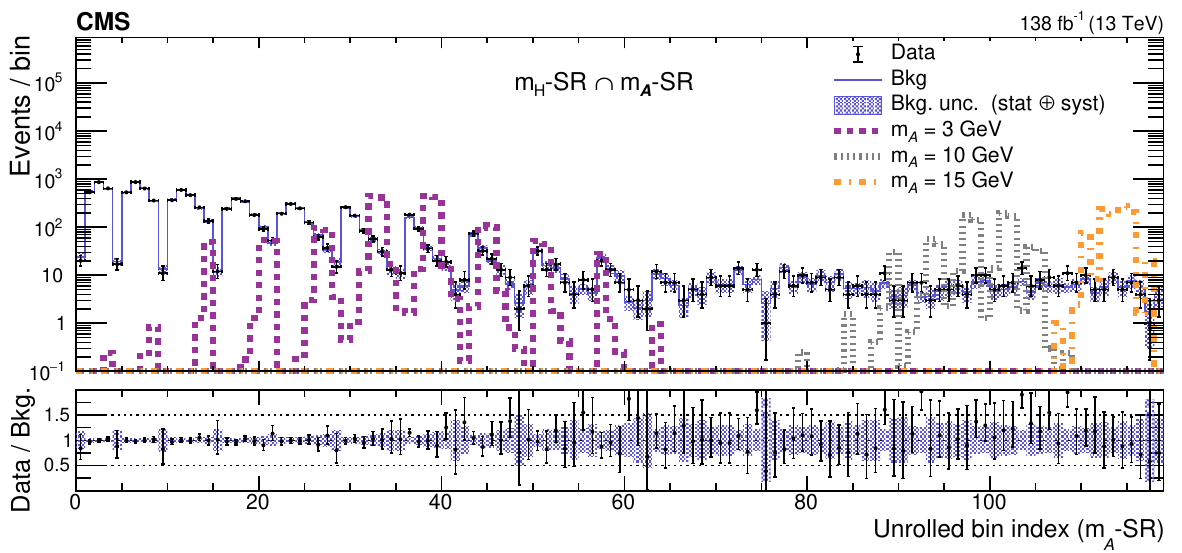}\\
\includegraphics[width=.45\textwidth]{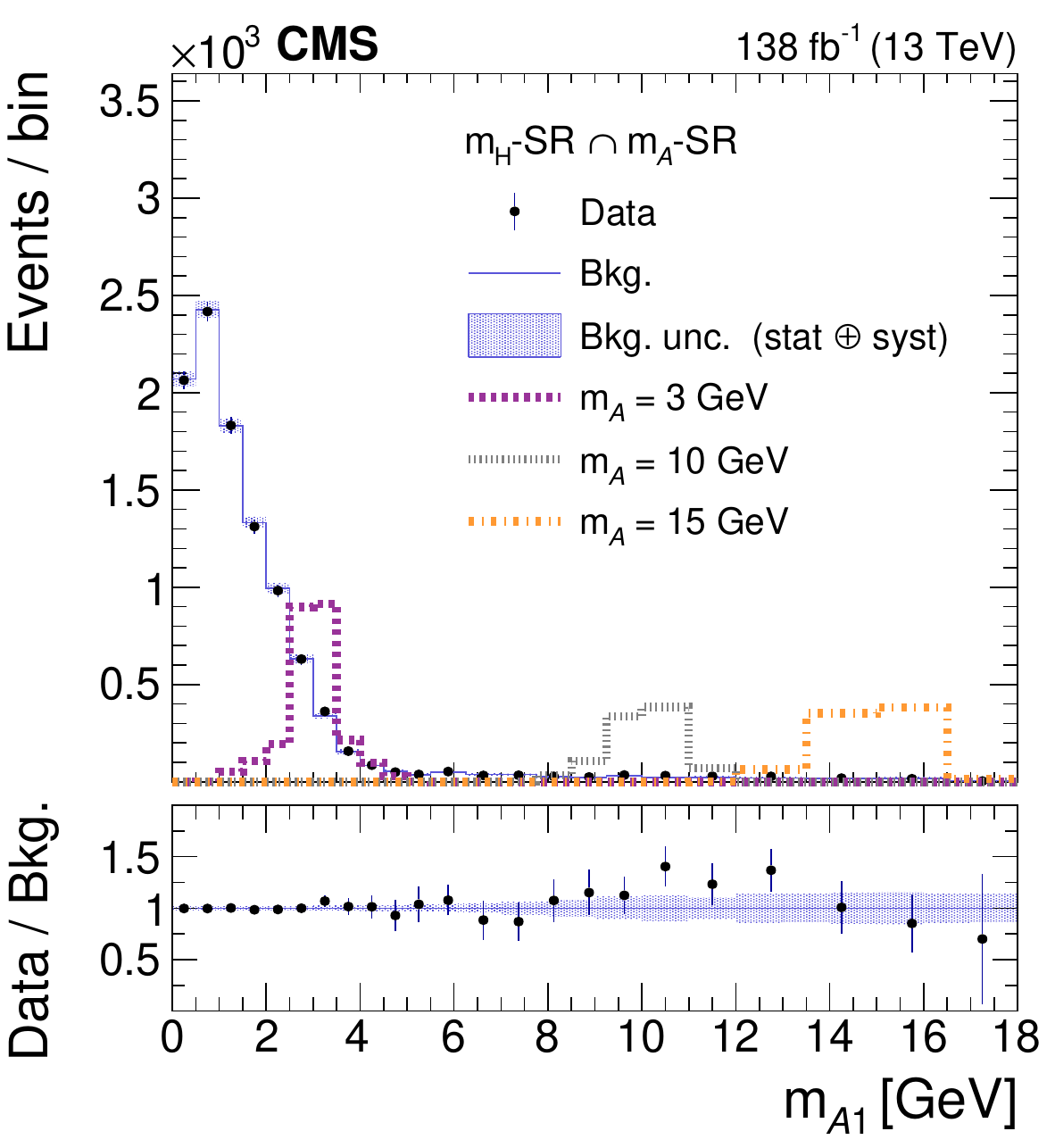}
\includegraphics[width=.45\textwidth]{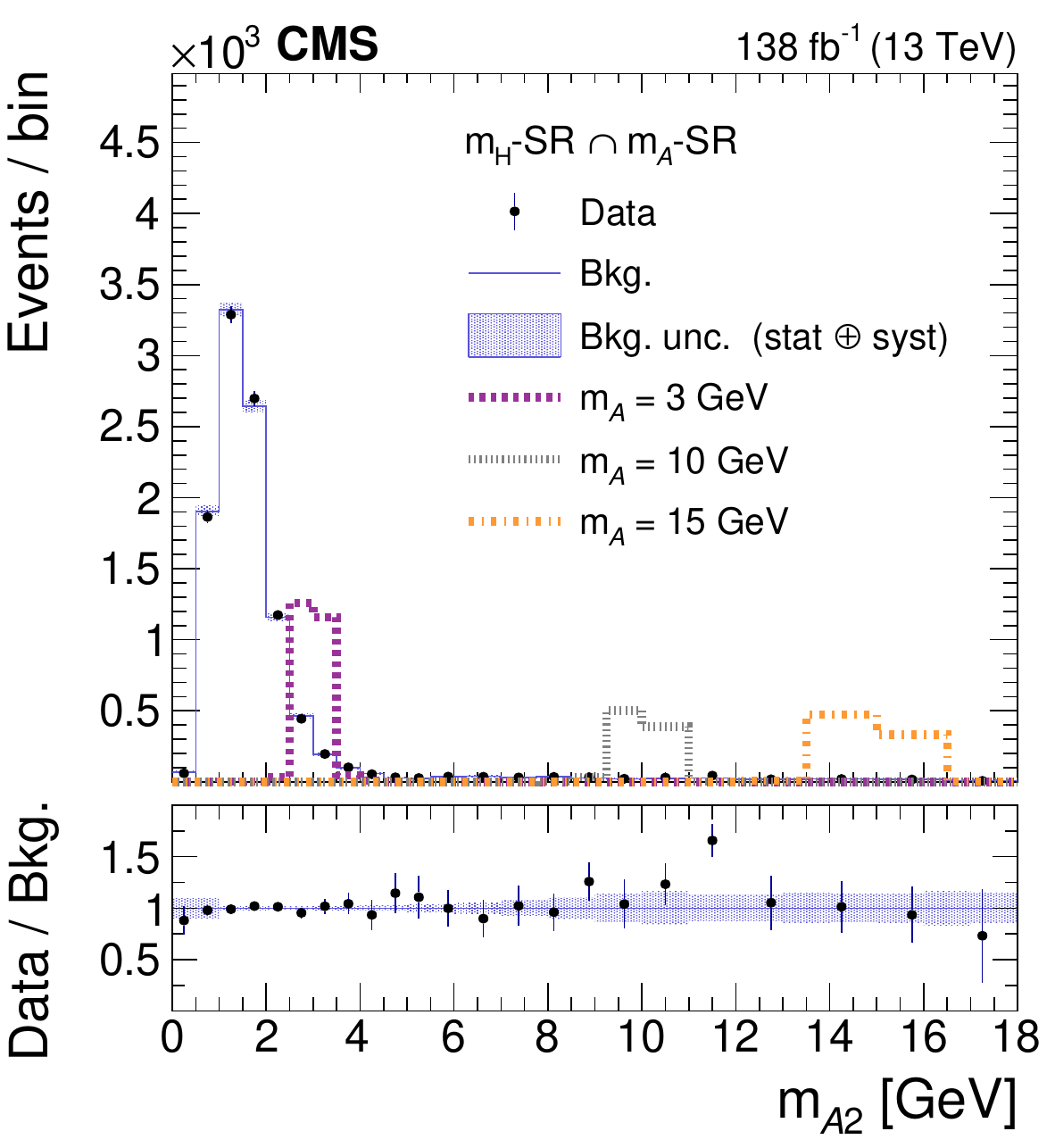}
\caption{The \mma spectra in the final signal region. Upper plot: The unrolled \mma distribution made by scanning along bins of increasing \mar at fixed \mam before incrementing in \mam. Only the bins in the \masr region are included, with the x-axis corresponding to the unrolled bin index of the selected bins, listed sequentially. 
Lower plot: 1D projections on the \mam (left) and \mar (right) axes of the \mma distribution. The data distributions (black points) are plotted against the total predicted background distributions (blue curves) after fitting to the data. The statistical plus systematic uncertainties in the background distribution are plotted as the blue band. The corresponding distributions of simulated \hag events for $\ma = 3$ (purple curve), 10 (gray curve), and 15\GeV (orange curve) are also overlaid on top. They are each normalized to the value of the expected upper limit to the signal cross section times 50. The lower panels of each plot show the ratio of the observed data over the predicted background as the black points, with the error bars representing the statistical uncertainties in the former. The ratio of the statistical plus systematic uncertainties in the background over the background prediction is shown as the blue band.}
\label{fig:postfit}
\end{figure}

Figure~\ref{fig:postfit} shows the \mma distribution spectra in the final SR. The upper plot displays the unrolled distribution of the \mma distribution. The unrolling begins at the lower-left corner of the 2D histogram and scans vertically in \mar at fixed \mam, then steps to the next \mam column. Bins outside the \masr region are skipped, with the x-axis indexing the retained bins in this scanning order.
The lower plots show the 1D projections of the \mma distribution onto the \mam (left) and \mar (right) axes. The unrolled 2D distribution is chosen as the final discriminant, since the 2D distribution provides better sensitivity than any individual 1D projection.  In each of the plots, the background model (in blue), after fitting to the data (in black), is shown overlaid with the signal model distributions for \ma = 3, 10, and 15\GeV. Good agreement between data and background prediction is observed. No statistically significant excess above the SM expectation is found in the explored mass range. 

\begin{figure}[htbp!]
\centering{
\includegraphics[width=0.55\textwidth]{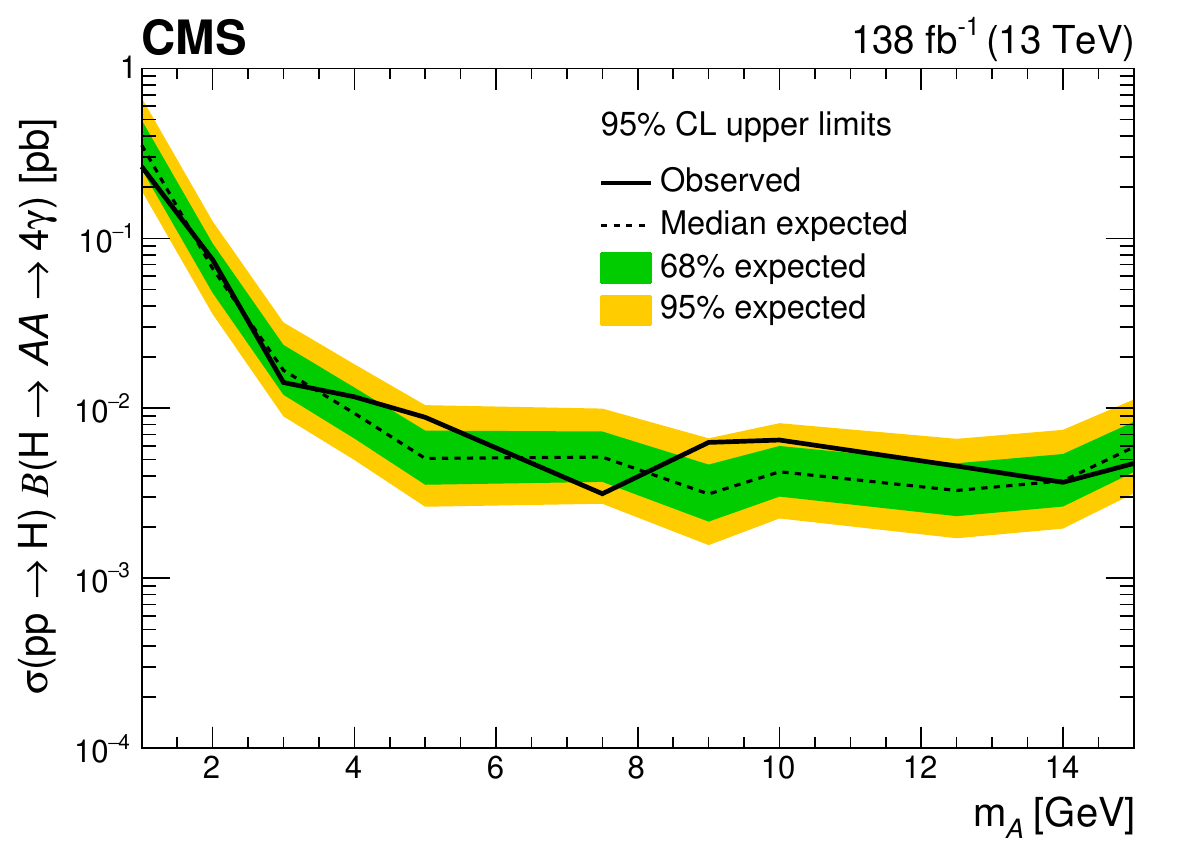} }
\caption{Observed (solid black line) and median expected (dashed black line) upper limits at 95\% \CL on the signal cross section times branching fraction $\xsh\,\Br(\hag)$ at various \ma mass points. The 68 and 95\% confidence intervals around the median expected limit are shown as the green and yellow bands, respectively. }
\label{fig:limit}
\end{figure}

The observed upper limits on the signal cross section times branching fraction \xsh\,\Br~(\hag) at 95\% \CL are plotted with a solid black line, with the median expected limit in a dashed black line, along with the 68 and 95\% \CL intervals in green and yellow bands, respectively, in Fig.~\ref{fig:limit}.
The 95\% \CL observed (expected) limits range from a value of  $0.264\,(0.354)\pb$ for $\ma = 1\GeV$ to $0.005\,(0.006)\pb$ for $\ma = 15\GeV$. 
The expected and observed limits are compatible across the full mass range.

\begin{figure}[htbp!]
\centering{
\includegraphics[width=0.55\textwidth]{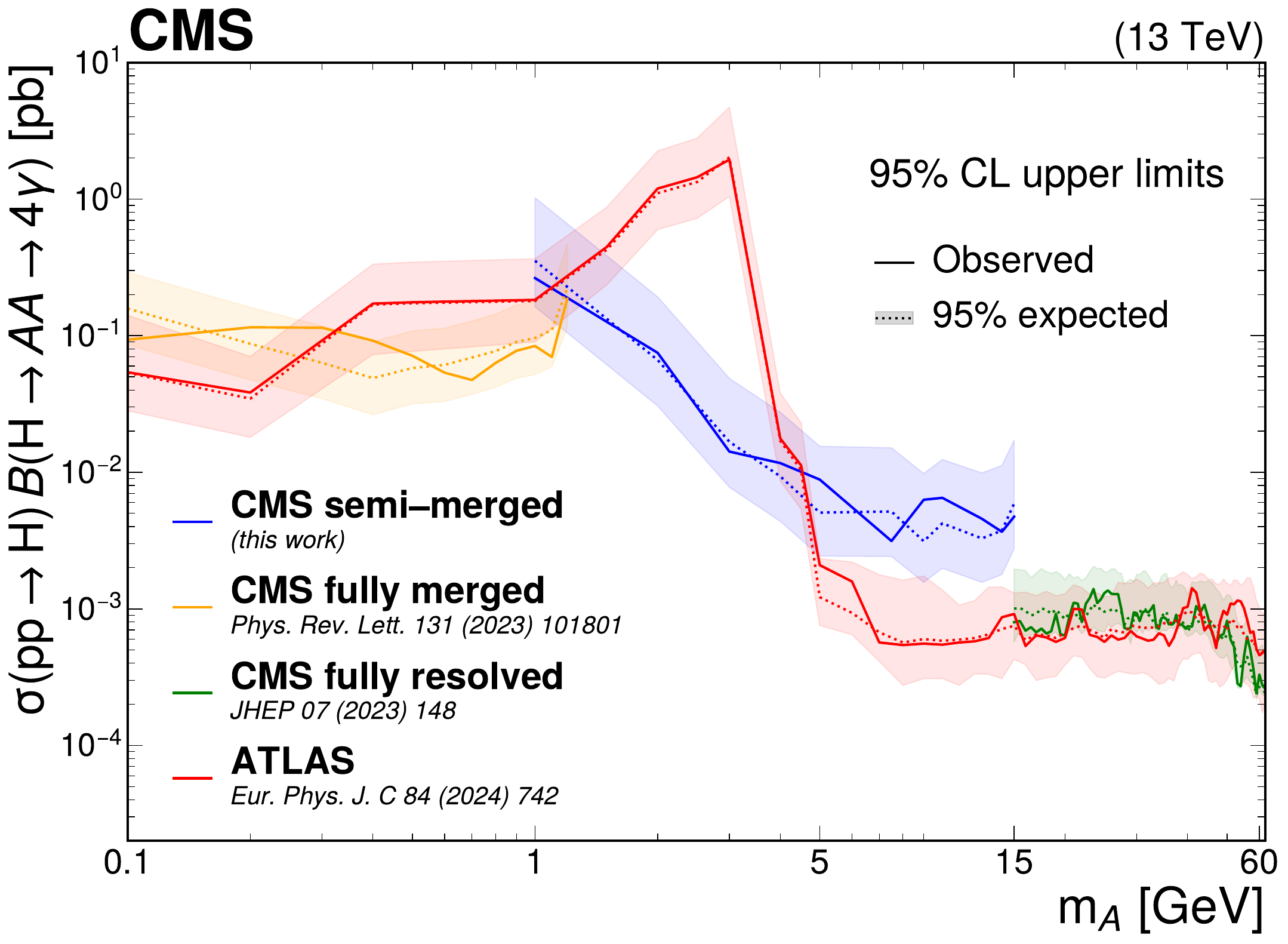} }
\caption{Observed (solid line) and expected (dashed line) upper limits, including the 95\% \CL interval around the expected limit, from this search (CMS semi-merged) compared to the previous CMS and ATLAS searches~\cite{CMS:Merged, CMS:Resolved, ATLAS:htoaato4g2023}, over the full \ma range of 0.1 to 60\GeV. The limits are shown in terms of $\xsh\,\Br(\hag)$.} 
\label{fig:limit-combined}
\end{figure}

Figure~\ref{fig:limit-combined} presents a comparison of the observed and expected upper limits from previous CMS and ATLAS searches with those obtained in this search. The largest improvement in our search compared to the ATLAS search is achieved at a mass of $\ma = 3\GeV$, where the observed limit is 100 times more stringent than the previous ATLAS result~\cite{ATLAS:htoaato4g2023}. The ATLAS Collaboration attributes their reduced sensitivity in this region to the transition between the merged and resolved photon categories, where the merged category has significantly larger background contributions.
The sensitivity of our search is reduced at the lowest mass compared to the previous CMS search, since the fully merged photon signature still dominates there, while at the highest masses it is reduced because of the small number of events, the use of coarser binning, and the large background uncertainty. This search with the semi-merged photon signatures thus enables the extraction of stringent limits across the entire mass spectrum, bridging the gap in sensitivity between the two previous CMS searches, and is also the first measurement from the CMS experiment in the mass range $\ma = 1$--15\GeV.
\newpage

\section{Summary}
\label{sec:Summary}

A search for the exotic decay of the Higgs boson of the form \hag in events with three photon-like objects has been performed using proton-proton collision data collected by the CMS experiment at $\sqrt{s} = 13\TeV$ corresponding to an integrated luminosity of 138\fbinv. One of the hypothetical particles, \Pa, is assumed to decay promptly to one semi-merged diphoton candidate reconstructed as a single photon-like object, while the other \Pa decays into a pair of resolved
photons. 
No excess above the estimated background is found. Upper limits are set on the product of the Higgs boson production cross section and branching fraction, $\xsh\,\Br(\hag)$, ranging from 0.264 to 0.005\pb  at 95\% confidence level for masses of \Pa in the range 1--15\GeV. These limits are the most stringent to date in the 1--5\GeV mass range.

\begin{acknowledgments}
We congratulate our colleagues in the CERN accelerator departments for the excellent performance of the LHC and thank the technical and administrative staffs at CERN and at other CMS institutes for their contributions to the success of the CMS effort. In addition, we gratefully acknowledge the computing centers and personnel of the Worldwide LHC Computing Grid and other centers for delivering so effectively the computing infrastructure essential to our analyses. Finally, we acknowledge the enduring support for the construction and operation of the LHC, the CMS detector, and the supporting computing infrastructure provided by the following funding agencies: SC (Armenia), BMBWF and FWF (Austria); FNRS and FWO (Belgium); CNPq, CAPES, FAPERJ, FAPERGS, and FAPESP (Brazil); MES and BNSF (Bulgaria); CERN; CAS, MoST, and NSFC (China); MINCIENCIAS (Colombia); MSES and CSF (Croatia); RIF (Cyprus); SENESCYT (Ecuador); ERC PRG, TARISTU24-TK10 and MoER TK202 (Estonia); Academy of Finland, MEC, and HIP (Finland); CEA and CNRS/IN2P3 (France); SRNSF (Georgia); BMFTR, DFG, and HGF (Germany); GSRI (Greece); NKFIH (Hungary); DAE and DST (India); IPM (Iran); SFI (Ireland); INFN (Italy); MSIT and NRF (Republic of Korea); MES (Latvia); LMTLT (Lithuania); MOE and UM (Malaysia); BUAP, CINVESTAV, CONACYT, LNS, SEP, and UASLP-FAI (Mexico); MOS (Montenegro); MBIE (New Zealand); PAEC (Pakistan); MES, NSC, and NAWA (Poland); FCT (Portugal); MESTD (Serbia); MICIU/AEI and PCTI (Spain); MOSTR (Sri Lanka); Swiss Funding Agencies (Switzerland); MST (Taipei); MHESI (Thailand); TUBITAK and TENMAK (T\"{u}rkiye); NASU (Ukraine); STFC (United Kingdom); DOE and NSF (USA).

\hyphenation{Rachada-pisek} Individuals have received support from the Marie-Curie program and the European Research Council and Horizon 2020 Grant, contract Nos.\ 675440, 724704, 752730, 758316, 765710, 824093, 101115353, 101002207, 101001205, and COST Action CA16108 (European Union); the Leventis Foundation; the Alfred P.\ Sloan Foundation; the Alexander von Humboldt Foundation; the Science Committee, project no. 22rl-037 (Armenia); the Fonds pour la Formation \`a la Recherche dans l'Industrie et dans l'Agriculture (FRIA) and Fonds voor Wetenschappelijk Onderzoek contract No. 1228724N (Belgium); the Beijing Municipal Science \& Technology Commission, No. Z191100007219010, the Fundamental Research Funds for the Central Universities, the Ministry of Science and Technology of China under Grant No. 2023YFA1605804, the Natural Science Foundation of China under Grant No. 12061141002, 12535004, and USTC Research Funds of the Double First-Class Initiative No.\ YD2030002017 (China); the Ministry of Education, Youth and Sports (MEYS) of the Czech Republic; the Shota Rustaveli National Science Foundation, grant FR-22-985 (Georgia); the Deutsche Forschungsgemeinschaft (DFG), among others, under Germany's Excellence Strategy -- EXC 2121 ``Quantum Universe" -- 390833306, and under project number 400140256 - GRK2497; the Hellenic Foundation for Research and Innovation (HFRI), Project Number 2288 (Greece); the Hungarian Academy of Sciences, the New National Excellence Program - \'UNKP, the NKFIH research grants K 131991, K 133046, K 138136, K 143460, K 143477, K 146913, K 146914, K 147048, 2020-2.2.1-ED-2021-00181, TKP2021-NKTA-64, and 2025-1.1.5-NEMZ\_KI-2025-00004 (Hungary); the Council of Science and Industrial Research, India; ICSC -- National Research Center for High Performance Computing, Big Data and Quantum Computing, FAIR -- Future Artificial Intelligence Research, and CUP I53D23001070006 (Mission 4 Component 1), funded by the NextGenerationEU program (Italy); the Latvian Council of Science; the Ministry of Education and Science, project no. 2022/WK/14, and the National Science Center, contracts Opus 2021/41/B/ST2/01369, 2021/43/B/ST2/01552, 2023/49/B/ST2/03273, and the NAWA contract BPN/PPO/2021/1/00011 (Poland); the Funda\c{c}\~ao para a Ci\^encia e a Tecnologia, grant CEECIND/01334/2018 (Portugal); the National Priorities Research Program by Qatar National Research Fund; MICIU/AEI/10.13039/501100011033, ERDF/EU, ``European Union NextGenerationEU/PRTR", and Programa Severo Ochoa del Principado de Asturias (Spain); the Chulalongkorn Academic into Its 2nd Century Project Advancement Project, the National Science, Research and Innovation Fund program IND\_FF\_68\_369\_2300\_097, and the Program Management Unit for Human Resources \& Institutional Development, Research and Innovation, grant B39G680009 (Thailand); the Kavli Foundation; the Nvidia Corporation; the SuperMicro Corporation; the Welch Foundation, contract C-1845; and the Weston Havens Foundation (USA).
\end{acknowledgments}

\bibliography{auto_generated} 

\providecommand{\href}[2]{#2}\begingroup\raggedright\begin{thebibliography}{10}%
\makeatletter
\providecommand{\hrefCMSnoop }[0]{\@secondoftwo}%
\makeatother
\providecommand{\doi}{\texttt{doi:}\begingroup \urlstyle{tt}\Url}

\bibitem{LHCpaper}
\hrefCMSnoop {}{L.~Evans and P.~Bryant, ``{LHC} machine'',} \textit{ JINST}
  \textbf{ 3} (2008) S08001,
  \href{http://dx.doi.org/10.1088/1748-0221/3/08/S08001}{\doi{10.1088/1748-0221/3/08/S08001}}.

\bibitem{ATLAS:2012yve}
\hrefCMSnoop {}{{ATLAS Collaboration}, ``Observation of a new particle in the
  search for the standard model {Higgs} boson with the {ATLAS} detector at the
  {LHC}'',} \textit{ Phys. Lett. B} \textbf{ 716} (2012) 1,
  \href{http://dx.doi.org/10.1016/j.physletb.2012.08.020}{\doi{10.1016/j.physletb.2012.08.020}},
  \href{http://www.arXiv.org/abs/1207.7214}{\texttt{arXiv:1207.7214}}.

\bibitem{CMS:2012qbp}
\hrefCMSnoop {}{{CMS Collaboration}, ``Observation of a new boson at a mass of
  125 {GeV} with the {CMS} experiment at the {LHC}'',} \textit{ Phys. Lett. B}
  \textbf{ 716} (2012) 30,
  \href{http://dx.doi.org/10.1016/j.physletb.2012.08.021}{\doi{10.1016/j.physletb.2012.08.021}},
  \href{http://www.arXiv.org/abs/1207.7235}{\texttt{arXiv:1207.7235}}.

\bibitem{CMS:higgslong2013}
\hrefCMSnoop {}{{CMS Collaboration}, ``Observation of a new boson with mass
  near 125 {GeV} in pp collisions at $\sqrt{s} = 7$ and 8 {TeV}'',} \textit{
  JHEP} \textbf{ 06} (2013) 081,
  \href{http://dx.doi.org/10.1007/JHEP06(2013)081}{\doi{10.1007/JHEP06(2013)081}},
  \href{http://www.arXiv.org/abs/1303.4571}{\texttt{arXiv:1303.4571}}.

\bibitem{CMS:2026combinedHiggsmeasurement}
\hrefCMSnoop {}{{CMS Collaboration}, ``Combined measurements and
  interpretations of {Higgs} boson production and decay at $\sqrt{s}=13$
  {TeV}'',} 2026.
  \href{http://www.arXiv.org/abs/2602.18611}{\texttt{arXiv:2602.18611}}.
  Submitted to \emph{ROPP}.

\bibitem{hexo}
D.~Curtin\hrefCMSnoop {}{ { et~al.}, ``Exotic decays of the 125 {GeV} {H}iggs
  boson'',} \textit{ Phys. Rev. D} \textbf{ 90} (2014) 075004,
  \href{http://dx.doi.org/10.1103/PhysRevD.90.075004}{\doi{10.1103/PhysRevD.90.075004}},
  \href{http://www.arXiv.org/abs/1312.4992}{\texttt{arXiv:1312.4992}}.

\bibitem{hexo2}
\hrefCMSnoop {}{D.~Curtin, R.~Essig, S.~Gori, and J.~Shelton, ``Illuminating
  dark photons with high-energy colliders'',} \textit{ JHEP} \textbf{ 02}
  (2015) 157,
  \href{http://dx.doi.org/10.1007/JHEP02(2015)157}{\doi{10.1007/JHEP02(2015)157}},
  \href{http://www.arXiv.org/abs/1412.0018}{\texttt{arXiv:1412.0018}}.

\bibitem{axiondmreview}
\hrefCMSnoop {}{F.~Chadha-Day, J.~Ellis, and D.~J.~E. Marsh, ``Axion dark
  matter: {What} is it and why now?'',} \textit{ Sci. Adv.} \textbf{ 8} (2022)
  abj3618,
  \href{http://dx.doi.org/10.1126/sciadv.abj3618}{\doi{10.1126/sciadv.abj3618}},
  \href{http://www.arXiv.org/abs/2105.01406}{\texttt{arXiv:2105.01406}}.

\bibitem{pqtheory}
\hrefCMSnoop {}{R.~D. Peccei and H.~R. Quinn, ``{CP} conservation in the
  presence of instantons'',} \textit{ Phys. Rev. Lett.} \textbf{ 38} (1977)
  1440,
  \href{http://dx.doi.org/10.1103/PhysRevLett.38.1440}{\doi{10.1103/PhysRevLett.38.1440}}.

\bibitem{axioncollider}
\hrefCMSnoop {}{M.~Bauer, M.~Neubert, and A.~Thamm, ``Collider probes of
  axion-like particles'',} \textit{ JHEP} \textbf{ 12} (2017) 044,
  \href{http://dx.doi.org/10.1007/JHEP12(2017)044}{\doi{10.1007/JHEP12(2017)044}},
  \href{http://www.arXiv.org/abs/1708.00443}{\texttt{arXiv:1708.00443}}.

\bibitem{axionreview}
\hrefCMSnoop {}{R.~D. Peccei, ``The strong {CP} problem and axions'',} in
  \textit{ Axions}, M.~Kuster, G.~Raffelt, and B.~Beltran, eds.
\newblock Springer Berlin Heidelberg, 2008.
\newblock
  \href{http://www.arXiv.org/abs/hep-ph/0607268}{\texttt{arXiv:hep-ph/0607268}}.
\newblock
  \href{http://dx.doi.org/10.1007/978-3-540-73518-2_1}{\doi{10.1007/978-3-540-73518-2_1}}.

\bibitem{CMS:2024uru}
\hrefCMSnoop {}{{CMS Collaboration}, ``Search for exotic decays of the {Higgs}
  boson to a pair of pseudoscalars in the $\mu\mu$bb and $\tau\tau$bb final
  states'',} \textit{ Eur. Phys. J. C} \textbf{ 84} (2024) 493,
  \href{http://dx.doi.org/10.1140/epjc/s10052-024-12727-4}{\doi{10.1140/epjc/s10052-024-12727-4}},
  \href{http://www.arXiv.org/abs/2402.13358}{\texttt{arXiv:2402.13358}}.

\bibitem{CMS:2024zfv}
\hrefCMSnoop {}{{CMS Collaboration}, ``Search for the decay of the {Higgs}
  boson to a pair of light pseudoscalar bosons in the final state with four
  bottom quarks in proton-proton collisions at $\sqrt{s} = 13$ {TeV}'',}
  \textit{ JHEP} \textbf{ 06} (2024) 097,
  \href{http://dx.doi.org/10.1007/JHEP06(2024)097}{\doi{10.1007/JHEP06(2024)097}},
  \href{http://www.arXiv.org/abs/2403.10341}{\texttt{arXiv:2403.10341}}.

\bibitem{CMS:2025hjt}
\hrefCMSnoop {}{{CMS Collaboration}, ``Search for light pseudoscalar boson
  pairs produced from {Higgs} boson decays using the 4$\tau$ and 2$\mu$2$\tau$
  final states in proton-proton collisions at $\sqrt{s} = 13$ {TeV}'',} 2025.
  \href{http://www.arXiv.org/abs/2508.06947}{\texttt{arXiv:2508.06947}}.
  Submitted to \emph{JHEP}.

\bibitem{ATLAS:2025qyn}
\hrefCMSnoop {}{{ATLAS Collaboration}, ``Search for {Higgs} boson exotic decays
  into {Lorentz}-boosted light bosons in the four-{\ensuremath{\tau}} final
  state at $\sqrt{s}=13$ {TeV} with the {ATLAS} detector'',} \textit{ Phys.
  Lett. B} \textbf{ 870} (2025) 139843,
  \href{http://dx.doi.org/10.1016/j.physletb.2025.139843}{\doi{10.1016/j.physletb.2025.139843}},
  \href{http://www.arXiv.org/abs/2503.05463}{\texttt{arXiv:2503.05463}}.

\bibitem{CMSlonglivedjets}
\hrefCMSnoop {}{{CMS Collaboration}, ``Search for light long-lived particles
  decaying to displaced jets in proton\textendash{}proton collisions at
  $\sqrt{s} = 13.6$ {TeV}'',} \textit{ Rept. Prog. Phys.} \textbf{ 88} (2025)
  037801,
  \href{http://dx.doi.org/10.1088/1361-6633/adaa13}{\doi{10.1088/1361-6633/adaa13}},
  \href{http://www.arXiv.org/abs/2409.10806}{\texttt{arXiv:2409.10806}}.

\bibitem{CMS:2018mumu}
\hrefCMSnoop {}{{CMS Collaboration}, ``A search for pair production of new
  light bosons decaying into muons in proton-proton collisions at 13 {TeV}'',}
  \textit{ Phys. Lett. B} \textbf{ 796} (2019) 131,
  \href{http://dx.doi.org/10.1016/j.physletb.2019.07.013}{\doi{10.1016/j.physletb.2019.07.013}},
  \href{http://www.arXiv.org/abs/1812.00380}{\texttt{arXiv:1812.00380}}.

\bibitem{CMS:2017yfk}
\hrefCMSnoop {}{{CMS Collaboration}, ``Particle-flow reconstruction and global
  event description with the {CMS} detector'',} \textit{ JINST} \textbf{ 12}
  (2017) P10003,
  \href{http://dx.doi.org/10.1088/1748-0221/12/10/P10003}{\doi{10.1088/1748-0221/12/10/P10003}},
\href{http://www.arXiv.org/abs/1706.04965}{\texttt{arXiv:1706.04965}}.

\bibitem{CMS:e2e-massreg}
\hrefCMSnoop {}{{CMS Collaboration}, ``Reconstruction of decays to merged
  photons using end-to-end deep learning with domain continuation in the {CMS}
  detector'',} \textit{ Phys. Rev. D} \textbf{ 108} (2023) 052002,
  \href{http://dx.doi.org/10.1103/PhysRevD.108.052002}{\doi{10.1103/PhysRevD.108.052002}},
  \href{http://www.arXiv.org/abs/2204.12313}{\texttt{arXiv:2204.12313}}.

\bibitem{CMS:Resolved}
\hrefCMSnoop {}{{CMS Collaboration}, ``Search for the exotic decay of the
  {Higgs} boson into two light pseudoscalars with four photons in the final
  state in proton-proton collisions at $\sqrt{s} = 13$ {TeV}'',} \textit{ JHEP}
  \textbf{ 07} (2023) 148,
  \href{http://dx.doi.org/10.1007/JHEP07(2023)148}{\doi{10.1007/JHEP07(2023)148}},
  \href{http://www.arXiv.org/abs/2208.01469}{\texttt{arXiv:2208.01469}}.

\bibitem{CMS:Merged}
\hrefCMSnoop {}{{CMS Collaboration}, ``Search for exotic {Higgs} boson decays
  $\rm{H} \to \mathcal{A}\mathcal{A} \to 4$\ensuremath{\gamma} with events
  containing two merged diphotons in proton-proton collisions at $\sqrt{s}$ =
  13 {TeV}'',} \textit{ Phys. Rev. Lett.} \textbf{ 131} (2023) 101801,
  \href{http://dx.doi.org/10.1103/PhysRevLett.131.101801}{\doi{10.1103/PhysRevLett.131.101801}},
  \href{http://www.arXiv.org/abs/2209.06197}{\texttt{arXiv:2209.06197}}.

\bibitem{ATLAS:htoaato4g2023}
\hrefCMSnoop {}{{ATLAS Collaboration}, ``Search for short- and long-lived
  axion-like particles in $\rm{H} \to a a \to 4$\ensuremath{\gamma} decays with
  the {ATLAS} experiment at the {LHC}'',} \textit{ Eur. Phys. J. C} \textbf{
  84} (2024) 742,
  \href{http://dx.doi.org/10.1140/epjc/s10052-024-12979-0}{\doi{10.1140/epjc/s10052-024-12979-0}},
  \href{http://www.arXiv.org/abs/2312.03306}{\texttt{arXiv:2312.03306}}.

\bibitem{CMS:2020uim}
\hrefCMSnoop {}{{CMS Collaboration}, ``Electron and photon reconstruction and
  identification with the {CMS} experiment at the {CERN} {LHC}'',} \textit{
  JINST} \textbf{ 16} (2021) P05014,
  \href{http://dx.doi.org/10.1088/1748-0221/16/05/P05014}{\doi{10.1088/1748-0221/16/05/P05014}},
  \href{http://www.arXiv.org/abs/2012.06888}{\texttt{arXiv:2012.06888}}.

\bibitem{hepdata}
\hrefCMSnoop {}{}{HEPD}ata record for this analysis, 2025.
\newblock
  \href{http://dx.doi.org/10.17182/hepdata.166011}{\doi{10.17182/hepdata.166011}}.

\bibitem{CMS:2008xjf}
\hrefCMSnoop {}{{CMS Collaboration}, ``The {CMS} experiment at the {CERN}
  {LHC}'',} \textit{ JINST} \textbf{ 3} (2008) S08004,
  \href{http://dx.doi.org/10.1088/1748-0221/3/08/S08004}{\doi{10.1088/1748-0221/3/08/S08004}}.

\bibitem{CMS:2023gfb}
\hrefCMSnoop {}{{CMS Collaboration}, ``Development of the {CMS} detector for
  the {CERN LHC Run 3}'',} \textit{ JINST} \textbf{ 19} (2024) P05064,
  \href{http://dx.doi.org/10.1088/1748-0221/19/05/P05064}{\doi{10.1088/1748-0221/19/05/P05064}},
  \href{http://www.arXiv.org/abs/2309.05466}{\texttt{arXiv:2309.05466}}.

\bibitem{CMS:2020cmk}
\hrefCMSnoop {}{{CMS Collaboration}, ``Performance of the {CMS} {Level-1}
  trigger in proton-proton collisions at $\sqrt{s} = 13$\,{TeV}'',} \textit{
  JINST} \textbf{ 15} (2020) P10017,
  \href{http://dx.doi.org/10.1088/1748-0221/15/10/P10017}{\doi{10.1088/1748-0221/15/10/P10017}},
  \href{http://www.arXiv.org/abs/2006.10165}{\texttt{arXiv:2006.10165}}.

\bibitem{CMS:2016ngn}
\hrefCMSnoop {}{{CMS Collaboration}, ``The {CMS} trigger system'',} \textit{
  JINST} \textbf{ 12} (2017) P01020,
  \href{http://dx.doi.org/10.1088/1748-0221/12/01/P01020}{\doi{10.1088/1748-0221/12/01/P01020}},
\href{http://www.arXiv.org/abs/1609.02366}{\texttt{arXiv:1609.02366}}.

\bibitem{CMS:2024aqx}
\hrefCMSnoop {}{{CMS Collaboration}, ``Performance of the {CMS} high-level
  trigger during {LHC} {Run} 2'',} \textit{ JINST} \textbf{ 19} (2024) P11021,
  \href{http://dx.doi.org/10.1088/1748-0221/19/11/P11021}{\doi{10.1088/1748-0221/19/11/P11021}},
  \href{http://www.arXiv.org/abs/2410.17038}{\texttt{arXiv:2410.17038}}.

\bibitem{CMS:2018rym}
\hrefCMSnoop {}{{CMS Collaboration}, ``Performance of the {CMS} muon detector
  and muon reconstruction with proton-proton collisions at $\sqrt{s}=$ 13
  {TeV}'',} \textit{ JINST} \textbf{ 13} (2018) P06015,
  \href{http://dx.doi.org/10.1088/1748-0221/13/06/P06015}{\doi{10.1088/1748-0221/13/06/P06015}},
  \href{http://www.arXiv.org/abs/1804.04528}{\texttt{arXiv:1804.04528}}.

\bibitem{CMS:2014pgm}
\hrefCMSnoop {}{{CMS Collaboration}, ``Description and performance of track and
  primary-vertex reconstruction with the {CMS} tracker'',} \textit{ JINST}
  \textbf{ 9} (2014) P10009,
  \href{http://dx.doi.org/10.1088/1748-0221/9/10/P10009}{\doi{10.1088/1748-0221/9/10/P10009}},
  \href{http://www.arXiv.org/abs/1405.6569}{\texttt{arXiv:1405.6569}}.

\bibitem{CMS:2018jrd}
\hrefCMSnoop {}{{CMS Collaboration}, ``Performance of reconstruction and
  identification of $\tau$ leptons decaying to hadrons and $\nu_\tau$ in pp
  collisions at $\sqrt{s}=$ 13 {TeV}'',} \textit{ JINST} \textbf{ 13} (2018)
  P10005,
  \href{http://dx.doi.org/10.1088/1748-0221/13/10/P10005}{\doi{10.1088/1748-0221/13/10/P10005}},
  \href{http://www.arXiv.org/abs/1809.02816}{\texttt{arXiv:1809.02816}}.

\bibitem{CMS:2016lmd}
\hrefCMSnoop {}{{CMS Collaboration}, ``Jet energy scale and resolution in the
  {CMS} experiment in pp collisions at 8 {TeV}'',} \textit{ JINST} \textbf{ 12}
  (2017) P02014,
  \href{http://dx.doi.org/10.1088/1748-0221/12/02/P02014}{\doi{10.1088/1748-0221/12/02/P02014}},
\href{http://www.arXiv.org/abs/1607.03663}{\texttt{arXiv:1607.03663}}.

\bibitem{CMS:2019ctu}
\hrefCMSnoop {}{{CMS Collaboration}, ``Performance of missing transverse
  momentum reconstruction in proton-proton collisions at $\sqrt{s} = 13$\,{TeV}
  using the {CMS} detector'',} \textit{ JINST} \textbf{ 14} (2019) P07004,
  \href{http://dx.doi.org/10.1088/1748-0221/14/07/P07004}{\doi{10.1088/1748-0221/14/07/P07004}},
\href{http://www.arXiv.org/abs/1903.06078}{\texttt{arXiv:1903.06078}}.

\bibitem{puppi}
\hrefCMSnoop {}{{CMS Collaboration}, ``Pileup mitigation at {CMS} in 13 {TeV}
  data'',} \textit{ JINST} \textbf{ 15} (2020) P09018,
  \href{http://dx.doi.org/10.1088/1748-0221/15/09/P09018}{\doi{10.1088/1748-0221/15/09/P09018}},
  \href{http://www.arXiv.org/abs/2003.00503}{\texttt{arXiv:2003.00503}}.

\bibitem{Madgraph:2014hca}
J.~Alwall\hrefCMSnoop {}{ { et~al.}, ``The automated computation of tree-level
  and next-to-leading order differential cross sections, and their matching to
  parton shower simulations'',} \textit{ JHEP} \textbf{ 07} (2014) 079,
  \href{http://dx.doi.org/10.1007/JHEP07(2014)079}{\doi{10.1007/JHEP07(2014)079}},
  \href{http://www.arXiv.org/abs/1405.0301}{\texttt{arXiv:1405.0301}}.

\bibitem{NNPDF:2017mvq}
\hrefCMSnoop {}{{NNPDF} Collaboration, ``Parton distributions from
  high-precision collider data'',} \textit{ Eur. Phys. J. C} \textbf{ 77}
  (2017) 663,
  \href{http://dx.doi.org/10.1140/epjc/s10052-017-5199-5}{\doi{10.1140/epjc/s10052-017-5199-5}},
  \href{http://www.arXiv.org/abs/1706.00428}{\texttt{arXiv:1706.00428}}.

\bibitem{Pythia8p2}
T.~Sj{\"o}strand\hrefCMSnoop {}{ { et~al.}, ``An introduction to {PYTHIA}
  8.2'',} \textit{ Comput. Phys. Commun.} \textbf{ 191} (2015) 159,
  \href{http://dx.doi.org/10.1016/j.cpc.2015.01.024}{\doi{10.1016/j.cpc.2015.01.024}},
  \href{http://www.arXiv.org/abs/1410.3012}{\texttt{arXiv:1410.3012}}.

\bibitem{CP5Tune:2019csb}
\hrefCMSnoop {}{{CMS Collaboration}, ``Extraction and validation of a new set
  of {CMS PYTHIA8} tunes from underlying-event measurements'',} \textit{ Eur.
  Phys. J. C} \textbf{ 80} (2020) 4,
  \href{http://dx.doi.org/10.1140/epjc/s10052-019-7499-4}{\doi{10.1140/epjc/s10052-019-7499-4}},
  \href{http://www.arXiv.org/abs/1903.12179}{\texttt{arXiv:1903.12179}}.

\bibitem{GEANT4:2002zbu}
\hrefCMSnoop {}{{GEANT4} Collaboration, ``{\GEANTfour}---a simulation
  toolkit'',} \textit{ Nucl. Instrum. Meth. A} \textbf{ 506} (2003) 250,
\href{http://dx.doi.org/10.1016/S0168-9002(03)01368-8}{\doi{10.1016/S0168-9002(03)01368-8}}.

\bibitem{CMS:2024yhz}
\hrefCMSnoop {}{{CMS Collaboration}, ``Search for a standard model-like {Higgs}
  boson in the mass range between 70 and 110 {GeV} in the diphoton final state
  in proton-proton collisions at $\sqrt{s}=13$ {TeV}'',} \textit{ Phys. Lett.
  B} \textbf{ 860} (2025) 139067,
  \href{http://dx.doi.org/10.1016/j.physletb.2024.139067}{\doi{10.1016/j.physletb.2024.139067}},
  \href{http://www.arXiv.org/abs/2405.18149}{\texttt{arXiv:2405.18149}}.

\bibitem{ATLASPF}
\hrefCMSnoop {}{{ATLAS Collaboration}, ``Jet reconstruction and performance
  using particle flow with the {ATLAS} detector'',} \textit{ Eur. Phys. J. C}
  \textbf{ 77} (2017) 466,
  \href{http://dx.doi.org/10.1140/epjc/s10052-017-5031-2}{\doi{10.1140/epjc/s10052-017-5031-2}},
  \href{http://www.arXiv.org/abs/1703.10485}{\texttt{arXiv:1703.10485}}.

\bibitem{hgg}
\hrefCMSnoop {}{{CMS Collaboration}, ``Observation of the diphoton decay of the
  {H}iggs boson and measurement of its properties'',} \textit{ Eur. Phys. J. C}
  \textbf{ 74} (2014) 3076,
  \href{http://dx.doi.org/10.1140/epjc/s10052-014-3076-z}{\doi{10.1140/epjc/s10052-014-3076-z}},
  \href{http://www.arXiv.org/abs/1407.0558}{\texttt{arXiv:1407.0558}}.

\bibitem{CMS:EGM-14-001}
\hrefCMSnoop {}{{CMS Collaboration}, ``Performance of photon reconstruction and
  identification with the {CMS} detector in proton-proton collisions at
  $\sqrt{s} = 8$ {TeV}'',} \textit{ JINST} \textbf{ 10} (2015) P08010,
  \href{http://dx.doi.org/10.1088/1748-0221/10/08/P08010}{\doi{10.1088/1748-0221/10/08/P08010}},
  \href{http://www.arXiv.org/abs/1502.02702}{\texttt{arXiv:1502.02702}}.

\bibitem{Ju_2020}
\hrefCMSnoop {}{X.~Ju and B.~Nachman, ``Supervised jet clustering with graph
  neural networks for {Lorentz} boosted bosons'',} \textit{ Phys. Rev. D}
  \textbf{ 102} (2020) 075014,
  \href{http://dx.doi.org/10.1103/PhysRevD.102.075014}{\doi{10.1103/PhysRevD.102.075014}},
  \href{http://www.arXiv.org/abs/2008.06064}{\texttt{arXiv:2008.06064}}.

\bibitem{ATLAS:2018wis}
\hrefCMSnoop {}{{ATLAS Collaboration}, ``Performance of top-quark and {W}-boson
  tagging with {ATLAS} in {Run} 2 of the {LHC}'',} \textit{ Eur. Phys. J. C}
  \textbf{ 79} (2019) 375,
  \href{http://dx.doi.org/10.1140/epjc/s10052-019-6847-8}{\doi{10.1140/epjc/s10052-019-6847-8}},
  \href{http://www.arXiv.org/abs/1808.07858}{\texttt{arXiv:1808.07858}}.

\bibitem{Komiske:2018cqr}
\hrefCMSnoop {}{P.~T. Komiske, E.~M. Metodiev, and J.~Thaler, ``Energy flow
  networks: {Deep} sets for particle jets'',} \textit{ JHEP} \textbf{ 01}
  (2019) 121,
  \href{http://dx.doi.org/10.1007/JHEP01(2019)121}{\doi{10.1007/JHEP01(2019)121}},
  \href{http://www.arXiv.org/abs/1810.05165}{\texttt{arXiv:1810.05165}}.

\bibitem{Ghosh:2018gyw}
S.~Ghosh\hrefCMSnoop {}{ { et~al.}, ``A simulation study to distinguish prompt
  photon from $\pi^0$ and beam halo in a granular calorimeter using deep
  networks'',} \textit{ JINST} \textbf{ 14} (2019) P01011,
  \href{http://dx.doi.org/10.1088/1748-0221/14/01/P01011}{\doi{10.1088/1748-0221/14/01/P01011}},
  \href{http://www.arXiv.org/abs/1808.03987}{\texttt{arXiv:1808.03987}}.

\bibitem{Uboldi_2022}
L.~Uboldi\hrefCMSnoop {}{ { et~al.}, ``Extracting low energy signals from raw
  {LArTPC} waveforms using deep learning techniques \textemdash{} {A} proof of
  concept'',} \textit{ Nucl. Instrum. Meth. A} \textbf{ 1028} (2022) 166371,
  \href{http://dx.doi.org/10.1016/j.nima.2022.166371}{\doi{10.1016/j.nima.2022.166371}},
  \href{http://www.arXiv.org/abs/2106.09911}{\texttt{arXiv:2106.09911}}.

\bibitem{Andrews_2020}
M.~Andrews\hrefCMSnoop {}{ { et~al.}, ``End-to-end jet classification of quarks
  and gluons with the {CMS Open Data}'',} \textit{ Nucl. Instrum. Meth. A}
  \textbf{ 977} (2020) 164304,
  \href{http://dx.doi.org/10.1016/j.nima.2020.164304}{\doi{10.1016/j.nima.2020.164304}},
  \href{http://www.arXiv.org/abs/1902.08276}{\texttt{arXiv:1902.08276}}.

\bibitem{Adams_2019}
\hrefCMSnoop {}{{MicroBooNE} Collaboration, ``Deep neural network for
  pixel-level electromagnetic particle identification in the {MicroBooNE}
  liquid argon time projection chamber'',} \textit{ Phys. Rev. D} \textbf{ 99}
  (2019) 092001,
  \href{http://dx.doi.org/10.1103/PhysRevD.99.092001}{\doi{10.1103/PhysRevD.99.092001}},
  \href{http://www.arXiv.org/abs/1808.07269}{\texttt{arXiv:1808.07269}}.

\bibitem{ExaTrkX:2020nyf}
{Exa.TrkX} Collaboration, \hrefCMSnoop {}{X.~Ju { et~al.}, ``Graph neural
  networks for particle reconstruction in high energy physics detectors'',} in
  \textit{ {Proc. 33rd Annual Conf. Neural Information Processing Systems}}.
\newblock 2020.
\newblock
  \href{http://www.arXiv.org/abs/2003.11603}{\texttt{arXiv:2003.11603}}.

\bibitem{Petrucciani_2015MINIAOD}
\hrefCMSnoop {}{{CMS Collaboration}, ``{Mini-AOD}: {A} new analysis data format
  for {CMS}'',} \textit{ J. Phys. Conf. Ser.} \textbf{ 664} (2015) 072052,
  \href{http://dx.doi.org/10.1088/1742-6596/664/7/072052}{\doi{10.1088/1742-6596/664/7/072052}},
  \href{http://www.arXiv.org/abs/1702.04685}{\texttt{arXiv:1702.04685}}.

\bibitem{Peruzzi_2020NANOAOD}
\hrefCMSnoop {}{{CMS Collaboration}, ``The {NanoAOD} event data format in
  {CMS}'',} \textit{ J. Phys. Conf. Ser.} \textbf{ 1525} (2020) 012038,
  \href{http://dx.doi.org/10.1088/1742-6596/1525/1/012038}{\doi{10.1088/1742-6596/1525/1/012038}}.

\bibitem{Gray:2020mcm}
\hrefCMSnoop {}{L.~Gray, T.~Klijnsma, and S.~Ghosh, ``A dynamic reduction
  network for point clouds'',} 2020.
  \href{http://www.arXiv.org/abs/2003.08013}{\texttt{arXiv:2003.08013}}.

\bibitem{Goodfellow-et-al-2016}
I.~Goodfellow, Y.~Bengio, and A.~Courville, ``Deep Learning''.
\newblock MIT Press, 2016.
\newblock \url{http://www.deeplearningbook.org}.

\bibitem{wang2019dynamicgraphcnnlearning}
Y.~Wang\hrefCMSnoop {}{ { et~al.}, ``Dynamic graph {CNN} for learning on point
  clouds'',} 2018.
  \href{http://www.arXiv.org/abs/1801.07829}{\texttt{arXiv:1801.07829}}.

\bibitem{CMS:WZpaper}
\hrefCMSnoop {}{{CMS Collaboration}, ``Measurement of the inclusive {W} and {Z}
  production cross sections in pp collisions at $\sqrt{s}=7$ {TeV}'',} \textit{
  JHEP} \textbf{ 10} (2011) 132,
  \href{http://dx.doi.org/10.1007/JHEP10(2011)132}{\doi{10.1007/JHEP10(2011)132}},
  \href{http://www.arXiv.org/abs/1107.4789}{\texttt{arXiv:1107.4789}}.

\bibitem{LHCHiggsCrossSectionWorkingGroup4:2016ypw}
\hrefCMSnoop {}{{LHC Higgs Cross Section Working Group}, ``Handbook of {LHC
  Higgs} cross sections: 4. {Deciphering} the nature of the {Higgs} sector'',}
  \textit{ CERN Yellow Rep. Monogr.} \textbf{ 2} (2017) 1,
  \href{http://dx.doi.org/10.23731/CYRM-2017-002}{\doi{10.23731/CYRM-2017-002}},
  \href{http://www.arXiv.org/abs/1610.07922}{\texttt{arXiv:1610.07922}}.

\bibitem{CMS:lumi2016}
\hrefCMSnoop {}{{CMS Collaboration}, ``Precision luminosity measurement in
  proton-proton collisions at $\sqrt{s} = 13$ {TeV} in 2015 and 2016 at
  {CMS}'',} \textit{ Eur. Phys. J. C} \textbf{ 81} (2021) 800,
  \href{http://dx.doi.org/10.1140/epjc/s10052-021-09538-2}{\doi{10.1140/epjc/s10052-021-09538-2}},
  \href{http://www.arXiv.org/abs/2104.01927}{\texttt{arXiv:2104.01927}}.

\bibitem{CMS-PAS-LUM-17-004}
\href {https://cds.cern.ch/record/2621960/}{{CMS Collaboration}, ``{CMS}
  luminosity measurement for the 2017 data-taking period at $\sqrt{s}$ = 13
  {TeV}'',} CMS Physics Analysis Summary CMS-PAS-LUM-17-004, 2018.

\bibitem{CMS-PAS-LUM-18-002}
\href {https://cds.cern.ch/record/2676164/}{{CMS Collaboration}, ``{CMS}
  luminosity measurement for the 2018 data-taking period at $\sqrt{s}$ = 13
  {TeV}'',} CMS Physics Analysis Summary CMS-PAS-LUM-18-002, 2019.

\bibitem{CLsMethod}
\href {https://cds.cern.ch/record/1379837}{{ATLAS and CMS Collaborations, and
  LHC Higgs Combination Group}, ``Procedure for the {LHC} {Higgs} boson search
  combination in {Summer} 2011'',} Technical Report CMS-NOTE-2011-005,
  ATL-PHYS-PUB-2011-11, 2011.

\bibitem{Junk-CL:1999kv}
\hrefCMSnoop {}{T.~Junk, ``Confidence level computation for combining searches
  with small statistics'',} \textit{ Nucl. Instrum. Meth. A} \textbf{ 434}
  (1999) 435,
  \href{http://dx.doi.org/10.1016/S0168-9002(99)00498-2}{\doi{10.1016/S0168-9002(99)00498-2}},
  \href{http://www.arXiv.org/abs/hep-ex/9902006}{\texttt{arXiv:hep-ex/9902006}}.

\bibitem{CLs:2002hq}
\hrefCMSnoop {}{A.~L. Read, ``Presentation of search results: {The} ${CL_s}$
  technique'',} \textit{ J. Phys. G} \textbf{ 28} (2002) 2693,
  \href{http://dx.doi.org/10.1088/0954-3899/28/10/313}{\doi{10.1088/0954-3899/28/10/313}}.

\bibitem{Cowan_2011}
\hrefCMSnoop {}{G.~Cowan, K.~Cranmer, E.~Gross, and O.~Vitells, ``Asymptotic
  formulae for likelihood-based tests of new physics'',} \textit{ Eur. Phys. J.
  C} \textbf{ 71} (2011) 1554,
  \href{http://dx.doi.org/10.1140/epjc/s10052-011-1554-0}{\doi{10.1140/epjc/s10052-011-1554-0}},
  \href{http://www.arXiv.org/abs/1007.1727}{\texttt{arXiv:1007.1727}}.
  [Erratum: \doi{10.1140/epjc/s10052-013-2501-z}].

\bibitem{CMS:COMBINE}
\hrefCMSnoop {}{{CMS Collaboration}, ``The {CMS} statistical analysis and
  combination tool: \textsc{Combine}'',} \textit{ Comput. Softw. Big Sci.}
  \textbf{ 8} (2024) 19,
  \href{http://dx.doi.org/10.1007/s41781-024-00121-4}{\doi{10.1007/s41781-024-00121-4}},
  \href{http://www.arXiv.org/abs/2404.06614}{\texttt{arXiv:2404.06614}}.

\end{thebibliography}\endgroup

\cleardoublepage \appendix\section{The CMS Collaboration \label{app:collab}}\begin{sloppypar}\hyphenpenalty=5000\widowpenalty=500\clubpenalty=5000\cmsinstitute{Yerevan Physics Institute, Yerevan, Armenia}
{\tolerance=6000
A.~Hayrapetyan, V.~Makarenko\cmsorcid{0000-0002-8406-8605}, A.~Tumasyan\cmsAuthorMark{1}\cmsorcid{0009-0000-0684-6742}
\par}
\cmsinstitute{Institut f\"{u}r Hochenergiephysik, Vienna, Austria}
{\tolerance=6000
W.~Adam\cmsorcid{0000-0001-9099-4341}, J.W.~Andrejkovic, L.~Benato\cmsorcid{0000-0001-5135-7489}, T.~Bergauer\cmsorcid{0000-0002-5786-0293}, M.~Dragicevic\cmsorcid{0000-0003-1967-6783}, C.~Giordano, P.S.~Hussain\cmsorcid{0000-0002-4825-5278}, M.~Jeitler\cmsAuthorMark{2}\cmsorcid{0000-0002-5141-9560}, N.~Krammer\cmsorcid{0000-0002-0548-0985}, A.~Li\cmsorcid{0000-0002-4547-116X}, D.~Liko\cmsorcid{0000-0002-3380-473X}, M.~Matthewman, I.~Mikulec\cmsorcid{0000-0003-0385-2746}, J.~Schieck\cmsAuthorMark{2}\cmsorcid{0000-0002-1058-8093}, R.~Sch\"{o}fbeck\cmsAuthorMark{2}\cmsorcid{0000-0002-2332-8784}, D.~Schwarz\cmsorcid{0000-0002-3821-7331}, M.~Shooshtari\cmsorcid{0009-0004-8882-4887}, M.~Sonawane\cmsorcid{0000-0003-0510-7010}, W.~Waltenberger\cmsorcid{0000-0002-6215-7228}, C.-E.~Wulz\cmsAuthorMark{2}\cmsorcid{0000-0001-9226-5812}
% ...
\par}
\vskip\cmsinstskip
\dag:~Deceased\\
$^{1}$Also at Yerevan State University, Yerevan, Armenia\\
$^{2}$Also at TU Wien, Vienna, Austria\\
% ...
\end{sloppypar}
\end{document}